\documentclass[journal]{IEEEtran}

\usepackage{cite}
\usepackage{graphicx}
\usepackage{amsmath, amsthm, amssymb}
\usepackage{algorithm,algorithmic}
\usepackage{enumerate}
\usepackage{balance, color}
\usepackage{multirow, verbatim}

\theoremstyle{plain} 
\newtheorem{defn}{Definition}
\newtheorem{thm}{Theorem}
\newtheorem{lem}[thm]{Lemma}

\theoremstyle{definition}

\newcommand{\vct}[1]{\ensuremath{\mathbf{#1}}}

\newcommand{\sign}{\ensuremath{\mathsf{Sign}}}
\newcommand{\mac}{\ensuremath{\mathsf{Mac}}}
\newcommand{\cbn}{\ensuremath{\mathsf{Combine}}}
\newcommand{\vrf}{\ensuremath{\mathsf{Verify}}}
\newcommand{\SMac}{\ensuremath{\mathsf{SpaceMac}}}
\newcommand{\ie}{{\em i.e.}}
\newcommand{\ea}{{\em et al.}}
\newcommand{\eg}{{\em e.g.}}

\newcommand{\beforeSpace}{\vspace*{-5pt}}
\newcommand{\afterSpace}{\vspace*{-5pt}}

\begin{document}
\title{Cooperative Defense Against Pollution Attacks in Network Coding Using SpaceMac}

\author{Anh~Le,
        Athina~Markopoulou \\
        University of California, Irvine
\thanks{
Email: {\tt \{athina, anh.le\}@uci.edu}. 
Tel: (+1) 949 824 1637. Mail: 4100 Calit2 Bldg, UC Irvine, Irvine, CA, 92697-2800.}
}

%

\maketitle
\pagestyle{empty}
\thispagestyle{empty}

\begin{abstract}

Intra-session network coding is known to be vulnerable to pollution attacks. In this work, first, we introduce a novel homomorphic MAC scheme called $\SMac$, which allows an intermediate node to verify if its received packets belong to a specific subspace, even if the subspace is expanding over time. Then, we use $\SMac$ as a building block to design a cooperative scheme that provides complete defense against pollution attacks: (i) it can detect polluted packets early at intermediate nodes and (ii) it can identify the exact location of all, even colluding, attackers, thus making it possible to eliminate them. Our scheme is cooperative: parents and children of any node cooperate to detect any corrupted packets sent by the node, and nodes in the network cooperate with a central controller to identify the exact location of all attackers.
We implement $\SMac$ in both C/C++ and Java as a library, and we make the library available online. Our evaluation on both a PC and an Android device shows that (i) $\SMac$'s algorithms can be computed quickly ($\sim 28$ $\mu$s in C/C++) and efficiently ($\sim 64$ KB in C/C++), and (ii) our cooperative defense scheme has both low computation ($\sim 190$ $\mu$s in C/C++) and low communication ($\sim$ 2\%) overhead, significantly less than other comparable state-of-the-art schemes.

\end{abstract} 


\section{Introduction}
\label{sec:intro}

\IEEEPARstart{T}{he} network coding paradigm advocates that intermediate nodes in a network should mix incoming packets instead of simply forwarding them, and receivers should decode to obtain the original packets. This idea, originally introduced by Ahlswede \ea~\cite{Ahlswede2000}, has been shown to bring benefits in terms of throughput and distributed operation of networks, and has received much attention. In this work, we consider networks that employ intra-session linear network coding.

An inherent weakness of network coding is that it is particularly vulnerable to pollution (a.k.a. Byzantine) attacks. Malicious nodes can inject corrupted packets into a network. These packets are combined and forwarded by downstream nodes, causing a large number of corrupted packets propagate in the network. This wastes resources of the network, such as bandwidth and CPU time, and eventually prevents the decoding of the original packets at the receivers. The detrimental effect of pollution attacks has been shown through both theoretical analysis \cite{Kim2010} as well as experimentation \cite{Dong2009, Gkantsidis2006}.

Proposed defense mechanisms against pollution attacks can be classified into three categories: error correction \cite{Cai2002, Jaggi2005, Koetter2007, Zhang2006}, attack detection \cite{Krohn2004, Gkantsidis2006, Agrawal2009, Li2010, Boneh2009,  Charles2006, Zhao2007, Ho2004, Kehdi2009, Yu2009, Zhang2011}, and locating attackers \cite{Jafarisiavoshani2008, Wang2010}.
In this paper, we are interested in the latter two approaches. In particular, we set out to design a complete defense system that can not only detect the pollution attack in a timely manner but also accurately locate and eliminate all pollution attackers. This allows for dealing with any attack early and at its root.
To the best of our knowledge, none of the existing defense mechanisms can provide this level of protection.

To this end, we first propose a novel homomorphic message authentication code (MAC) scheme for expanding spaces called $\mathsf{SpaceMac}$. $\SMac$ allows a node to verify if its received packets belong to a specific subspace, even if the subspace is expanding over time. We then design our novel cooperative defense system which includes both a detection scheme and a locating scheme, using $\mathsf{SpaceMac}$ as their building block. Our detection scheme relies on $\mathsf{SpaceMac}$ to force intermediate nodes to send only linear combinations of packets that they actually receive from their parents. Parents and children of any intermediate node cooperate to detect corrupted packets sent by the intermediate node. Our locating scheme uses $\mathsf{SpaceMac}$ to force nodes in the network to truthfully cooperate with a central controller so that the controller can exactly locate the pollution attackers. Finally, by leveraging multiple generations, our scheme is able to deal with a large number of colluding attackers.


The main contribution of this paper is twofold:
\begin{itemize}
\item {\bf The design and implementation of $\SMac$}: We describe the construction of $\SMac$ and provide a formal security proof for the construction. We implement $\SMac$ in both C/C++ and Java as a ready-to-use library. Our Java implementation is compatible with the current Android OS (Android 2.2 Froyo). We make the library available online \cite{SpaceMacLib}.
\item {\bf The design of a novel cooperative defense system based on $\SMac$}: To the best of our knowledge, our defense system is the first that meets all of the following requirements simultaneously: (i) it can provide timely in-network detection, (ii) it can exactly locate all pollution attackers, (iii) it can deal with a large number of colluding attackers, and (iv) it has low communication and computation overhead.
\end{itemize}

We have extensively evaluated the computation overhead of $\SMac$'s algorithms and both the computation and communication overhead of our defense scheme through real implementation in both C/C++ and Java, and on both a PC and an Android device (Samsung Captivate). Our evaluation results show that all three algorithms of $\SMac$ ($\mac$, $\cbn$, and $\vrf$) can be computed efficiently (requiring 64 KB of memory in C/C++ or 128 KB in Java) and also quickly on a PC ($<$ 28 $\mu$s in C/C++) and even on a smart phone ($<$ 2.3 ms). Evaluation results also demonstrate that when implementing our defense scheme, nodes in the network introduce very small computational delay (in the order of sub-millisecond on the PC and millisecond on the smart phone). Moreover, our defense scheme was shown to introduce very low communication overhead (2\%), significantly less than other comparable state-of-the-art schemes. Lastly, through a simulation in Python, we show that in a medium-size network of 50 nodes, our locating scheme can quickly locate all, even colluding attackers (20 attackers in about 1 second).

The rest of this paper is organized as follows. Section \ref{sec:related} discusses existing approaches to protect network coding against pollution attacks. Section \ref{sec:formulation} formulates the problem, describes the threat model, and discusses our design goals. Section \ref{sec:observation} presents our key observations and the overview of our approach. Section \ref{sec:construction} presents the construction of $\SMac$ and the formal security proof. Section \ref{sec:detection} describes our detection scheme. Section \ref{sec:locating} describes our locating scheme. Section \ref{sec:security} analyzes the security of our proposed schemes. Section \ref{sec:evaluation} presents our implementation and the evaluation results. Finally, we conclude in Section \ref{sec:conclusion}.

\section{Related Work}
\label{sec:related}
There are three main approaches in the literature to combat pollution attacks: error correction, attack detection, and locating attacks.
Below, we discuss each one in detail.

\subsection{Error Correction}
One of the earliest work on error correction for network coding is by Cai and Yeung \cite{Cai2002}. The study in \cite{Cai2002} introduces network error-correcting codes as a generalization of the traditional error correction codes. In a related study by Zhang \cite{Zhang2006}, the minimum rank of a network error correction code is defined; this concept is analogous to the minimum distance in classical coding theory. Based on this concept, a network error correction codes similar to an ordinary linear network single source multicast code is designed. Jaggi \ea~\cite{Jaggi2007} consider packets from an attacker as an additional source and add redundancy at the source so that the receivers can decode both sources: the original source and adversary's source. In \cite{Koetter2007}, Koetter and Kschischang proposed a coding metric on subspaces and a minimum distance decoder, which give rise to codes capable of correcting certain combinations of errors and erasures.

These information theoretic approaches, which aim at correcting errors at the receivers, offer only limited security against restricted types of adversaries. These approaches assume that the adversaries can only corrupt a small number of edges and packets. Also, the amount of redundancy, which can also be considered as the communication overhead, typically increases proportional to the number of corrupted packets or adversaries. Furthermore, these approaches do not detect and drop corrupted packets, and thus are unable to prevent the corrupted packets from propagating in the network and using up resources.

In contrast, our defense scheme is able to provide timely detection of the attack, thereby allowing for early filtering of corrupted packets. More importantly, our approach can accurately locate the attackers to eliminate them from the network. Furthermore, our approach can deal with more powerful adversaries, \emph{i.e.}, adversaries who pollute arbitrary number of packets and even colluding adversaries. However, we make an assumption on the adversaries' computational power, which is typical of all approaches utilizing cryptographic primitives.

\subsection{Attack Detection}
We first describe approaches that
do not use homomorphic cryptography.
In \cite{Ho2004}, Ho \emph{et al.} show that randomized network coding can be extended to provide end-to-end attack detection, \ie, allow the receivers to detect any corrupted packet. The extension requires the source node to include in each source packet some additional hash blocks calculated from the source data blocks using polynomial functions. More recently, Kehdi and Li \cite{Kehdi2009} propose an in-network detection scheme which exploits subspace properties of network coding. In their scheme, intermediate nodes verify the integrity of a vector by checking if it belongs to the subspace spanned by the source vectors. Null keys, which are vectors orthogonal to all the combinations of the source vectors, are used for the verification. This scheme is not collusion resistant: multiple nodes can collude to infer the null keys and make benign nodes accept polluted vectors. Yu \ea~\cite{Yu2009} use simple XOR checksums and exploit probabilistic key pre-distribution to provide in-network detection. This scheme, however, has significant communication overhead due to the aggregation of authentication tags; moreover, the scheme is {\em c}-collusion resistant for some pre-determined constant {\em c}, \ie, the scheme becomes vulnerable when there are more than {\em c} colluding attackers. In \cite{Dong2009}, Dong \ea~ design a linear transformation checksums to be used with a time-based authentication scheme to provide in-network detection. This scheme requires time synchronization among nodes in the network and frequent public key verification (one per generation).

A significant number of homomorphic cryptographic primitives ranging from hashes, signatures, to MACs, has been designed specifically to combat pollution attacks in network coding. In \cite{Krohn2004}, Krohn \ea~proposed a homomorphic hash scheme for verification of rateless erasure codes. Gkantsidis and Rodriguez \cite{Gkantsidis2006} later propose probabilistic checking and cooperation mechanisms among nodes to reduce the computation overhead when using Krohn \ea 's scheme in peer-to-peer file distribution systems. Li \ea~ \cite{Li2006} also propose a hash-based scheme based on a trapdoor one-way permutation which can avoid the pre-distribution of the hash blocks.
In \cite{Zhao2007}, Zhao \ea~ propose a signature scheme where the source derives authentication information from a vector orthogonal to the source space. In \cite{Charles2006}, Charles \ea~ propose a signature scheme based on aggregate signatures. Recently, Boneh \ea~ presents a signature scheme built on bilinear maps \cite{Boneh2009}. All of the hash-based and signature-based approaches suffer from a common drawback: they require expensive computation at intermediate nodes, either modular exponentiation or bilinear map, which results in high latency.

Recently, two homomorphic MAC schemes are proposed by two groups of researchers: Agrawal and Boneh \cite{Agrawal2009}, and Li \ea~ \cite{Li2010}. The scheme in \cite{Agrawal2009} relies on cover free set systems for pre-distributing keys to provide in-network detection, and thus, only $c$-collusion resistant. This scheme is also susceptible to tag-pollution attacks, where malicious nodes tamper with some subset of tags of a packet. We discuss about this type of attack in detail in Section \ref{subsec:tag_pollution}. The scheme in \cite{Li2010} is collusion resistant (rather than $c$-collusion resistant) as well as resistant against tag-pollution attacks; however, it requires time synchronization among nodes in the network. Both schemes have low computation overhead since they only require simple addition and multiplication operations at intermediate nodes for both combining MAC tags and tag verification. In a more recent work \cite{Zhang2011}, Zhang \ea~introduce both a homomorphic MAC and a homomorphic signature scheme and propose a hybrid approach that uses both. This approach is not susceptible to tag-pollution attack but only $c$-collusion resistant. This approach also has lower computation overhead than signature-based approaches; however, the overhead is still significantly higher than the other two MAC-based approaches due to the expensive exponentiation operations required by the signature scheme.

Our $\SMac$ scheme is inspired by the MAC scheme in \cite{Agrawal2009}; however, our scheme allows intermediate nodes to sign -{\em expanding} over time- subspaces. This stands in stark contrast to \cite{Agrawal2009}, which allows for signing only fixed subspaces. Our scheme can be considered as a generalization of the scheme in \cite{Agrawal2009}. Detailed description and comparison are provided in Section \ref{sec:construction}. 
$\SMac$ was originally introduced for authenticating expanding subspaces and locating the attackers in our preliminary work \cite{Le2010}. In this paper, we show that the ability to authenticate expanding subspaces can also be utilized to provide timely in-network detection without requiring time synchronization as in \cite{Li2010}.


\subsection{Locating Attackers}
Compared to the other two categories, locating attackers has received less attention. An early work by Jafarisiavoshani \ea~ \cite{Jafarisiavoshani2008} leverages the subspace properties of randomized network coding to locate pollution attackers. The main observation is that packets sent by a node have to belong to the space spanned by source packets and also the space spanned by the packets the node receives from its parents. Using this observation, in a general network topology having a single attacker, the authors can locate the attacker with an uncertainty of at most two nodes; when there are multiple attackers, the uncertainty is within a set of nodes including the attackers and their parents and children. Our scheme builds on and significantly improves this work: we make it possible to pinpoint the exact location of the attackers, even in the case where there are multiple colluding attackers, thereby allowing for the removal of all attackers.

Recently, Wang \ea~ \cite{Wang2010} introduce a light-weight non-repudiation protocol ensuring that (i) a malicious node that injected a polluted packet cannot deny its behavior and (ii) a malicious node cannot disparage any innocent node. They build a defense scheme based on the protocol to identify malicious nodes. The scheme in \cite{Wang2010} performs flooding of multiple checksums of all the packets sent by the source to all the nodes, which incurs significant communication overhead. Moreover, because the success of the locating scheme relies on the successful reception of the checksums at every node, this scheme is vulnerable to colluding attackers. Finally, this scheme is unable to locate all the attackers. We use their non-repudiation protocol as a building block of our locating scheme. However, unlike the  scheme in \cite{Wang2010}, our scheme is able to locate all, even colluding attackers, without the need of checksums.

Finally, compared to our prior preliminary work \cite{Le2010}, where we first presented $\SMac$ and our locating scheme, this work is significantly improved and extended by the following three novel contributions: (i) we describe a novel construction of $\SMac$, whose algorithms are significantly more computational efficient than our previous construction; (ii) we describe a novel detection scheme built on $\SMac$ that can provide in-network detection with low overhead; and (iii) we implement the $\SMac$ library in both C/C++ and Java and we extensively evaluate both $\SMac$ and the proposed defense scheme on both a PC and an Android device.

\section{Problem Formulation}
\label{sec:formulation}

\begin{table}[tp]
\beforeSpace
\centering
\begin{tabular}{|l|l|}
\hline
{\bf Symbol} & {\bf Explanation}\\
\hline
$\vct{v}_i$ & source packet, formed by augmenting $\vct{\hat{v}}_i$ with coefficients\\
\hline
$n$ & the number of symbols carried in each packet $\vct{\hat{v}_i}$\\
\hline
$m$ & size of a generation\\
\hline
$\mathcal{P}_N, \mathcal{C}_N$ & the parent set and the child set of $N$\\
\hline
$S, \mathcal{R}, \mathcal{I}$ & source node, receiver set, intermediate node set\\
\hline
$\Pi^S$ & source space, spanned by source packets\\
\hline
$\Pi^N$ & sending space, spanned by packets sent by $N$\\
\hline
$\Pi_N^P, \Pi_N$ & received space of $N$ from $P$ and from all parents\\
\hline
\end{tabular}
\caption{\textnormal{Notation}}
\label{table:notation}
\afterSpace
\end{table}

In this section, we describe the notation that we use to express network operations in a multicast session with intra-session coding. In addition, we describe the threat model and the design goals of our defense system.

\subsection{Network Model and Operation}
We follow the notation used in \cite{Agrawal2009} and \cite{Jafarisiavoshani2008}. Consider a fixed, directed acyclic graph (DAG), denoted by $G=(\mathcal{V}, \mathcal{E})$. There is a single source node $S$ that multicasts packets to a set of receivers, denoted by $\mathcal{R}$. Denote the set of intermediate nodes as $\mathcal{I}$, \ie,  $\mathcal{I} = \mathcal{V} \setminus  \{ \mathcal{R} \cup \{S\} \}$. Nodes in $\mathcal{I}$ perform generation-based linear network coding.
A generation consists of $m$ packets, $\vct{\hat{v}}_1, \cdots, \vct{\hat{v}}_m$, in an $n$-dimensional linear space  $\mathbb{F}^{n}_{q}$, where $m, n$ and $q$ are fixed ahead of time and $q \gg 1$. The source augments every packet $\mathbf{\hat{v}}_i$ with $m$ additional symbols, which are the coefficients of $\vct{\hat{v}}_i$. The resulting packets, $\vct{v}_i$'s, called \emph{source packets}, have the following form:
\[\vct{v}_i = (\textrm{---}\vct{\hat{v}}_i\textrm{---}, \overbrace{\underbrace{0, \cdots, 0, 1}_i, 0, \cdots, 0}^m)\,\in \mathbb{F}^{n+m}_q\,.\]
Note that if a packet $\vct{y}$ is a linear combination of the packets $\vct{v}_i$'s, then the last $m$ symbols of packet $\vct{y}$ contain the global linear combination coefficients. The source $S$ sends the packets $\vct{v}_i$'s to the network. Denote the subspace spanned by packets $\vct{v}_i$'s by $\Pi^S \subseteq \mathbb{F}^{n+m}_q$. We refer to $\Pi^S$ as the {\em source space}.

We use $\mathcal{P}_N$ and $\mathcal{C}_N$ to denote the sets of parents and children of a node $N$, respectively. Each intermediate node $N$ receives from $\mathcal{P}_N$ some packets, which are linear combinations of the source packets. It then creates linear combinations of the received packets and sends them to its adjacent downstream nodes. We use $\Pi_N^{P} (t) \subseteq  \mathbb{F}^{n+m}_q$ to denote the space spanned by the packets received by node $N$ from a parent node $P$, $P \in \mathcal{P}_N$, up to time $t$. We further use $\Pi_N (t)$ to denote the space spanned by all the packets received by node $N$ from all its parents up to time $t$. We also denote the space spanned by all the packets sent by a node $N$ up to time $t$ as $\Pi^N(t)$. When there is no ambiguity, we omit the time index $t$. A receiver $R \in \mathcal{R}$ can successfully decode the original source packets using Gaussian elimination if its received space $\Pi_R$ equals the source space $\Pi^S$.

If all the nodes in the network are benign, then the space spanned by packets sent by $N$, $\Pi^N$, must be a subspace of the space spanned by the packets that $N$ receives, $\Pi_N$. This is a property of networks that implement random linear network coding. This observation was also made in \cite{Jafarisiavoshani2008}. Formally,
\begin{lem}\label{lem1}
If every node in the network is benign then for every node $N$, $\Pi^N(t) \subseteq \Pi_N(t)$.
\end{lem}
Also, observe that for any parent $P$ of $N$, both $\Pi^P_N$ and $\Pi_N$ expand over time. Formally, $\Pi^P_N(t_0) \subseteq \Pi^P_N(t_1)$ and $\Pi_N(t_0) \subseteq \Pi_N(t_1)$, for all $t_0 \leq t_1$

Furthermore, when all the intermediate nodes $N \in \mathcal{I}$ are benign, the incoming spaces of all the intermediate nodes and the receivers are subspaces of the source space.  Assume that there is a pollution attacker in the network. The attacker combines a subspace $\Pi^* \nsubseteq \Pi^S$ with its incoming space and sends the resulting packets to its children; as a result, the incoming spaces of these children are not subspaces of $\Pi^S$. Formally, for every node $N, N \in \mathcal{V} \setminus \{S\}$, its incoming space $\Pi_N^{P}$ from its parent $P$ can be decomposed as follows: $\Pi_N^{P} = \Pi^{S_P}_{N} \oplus \hat{\Pi}_N^{P}\,,$
where $\oplus$ denotes the direct sum of spaces, $\Pi ^{S_P}_{N} \overset{\text{def}}{=} \Pi^S \cap \Pi_N^{P}$, and $\hat{\Pi}_{N}^{P}$ contains packets not belonging to $\Pi^S$. Table \ref{table:notation} summarizes the notation used in this paper.

Following the framework in \cite{Jafarisiavoshani2008}, we define a \emph{polluted directed edge} as follows:
\begin{defn}\label{defn:polluted}
A directed edge is polluted if it transmits any packet which is not a  linear combination of the source packets.
\end{defn}
The following lemma, adapted from \cite{Jafarisiavoshani2008}, is a direct consequence of the definition:
\begin{lem}\label{lem2} A directed edge e$(P,N)$ is polluted if and only if $\Pi_N^P \nsubseteq \Pi^S\,$.
\end{lem}

\subsection{Threat Model}
We assume that both the source and the receivers are trustworthy but the intermediate nodes may be malicious. The case when the receivers are malicious is discussed separately in Section \ref{subsec:mal_receivers}. We assume that the network may have multiple pollution attackers. They may be located at an arbitrary set of intermediate nodes in the network.  Each attacker may inject corrupted packets, \ie, packets that do not belong to the source space, into a single or multiple downstream edges to pollute the network. They may also modify other data associated with the packets, such as, authentication tags. Successful modification of authentication tags constitutes an attack called ``tag pollution,'' which could be as devastating as a pollution attack \cite{Li2010}. Tag pollution attacks are discussed separately in Section \ref{subsec:tag_pollution}. We consider both cases where the pollution attackers launch their attacks independently or collude and coordinate their attacks. We further assume that the attackers are aware of our defense scheme, \ie, the construction and application of $\SMac$; however, similar to other cryptographic approaches, we assume that the attackers' running time is polynomial in the security parameter.

\begin{figure}[t!]
\beforeSpace
\centering
\includegraphics[width=8.5cm]{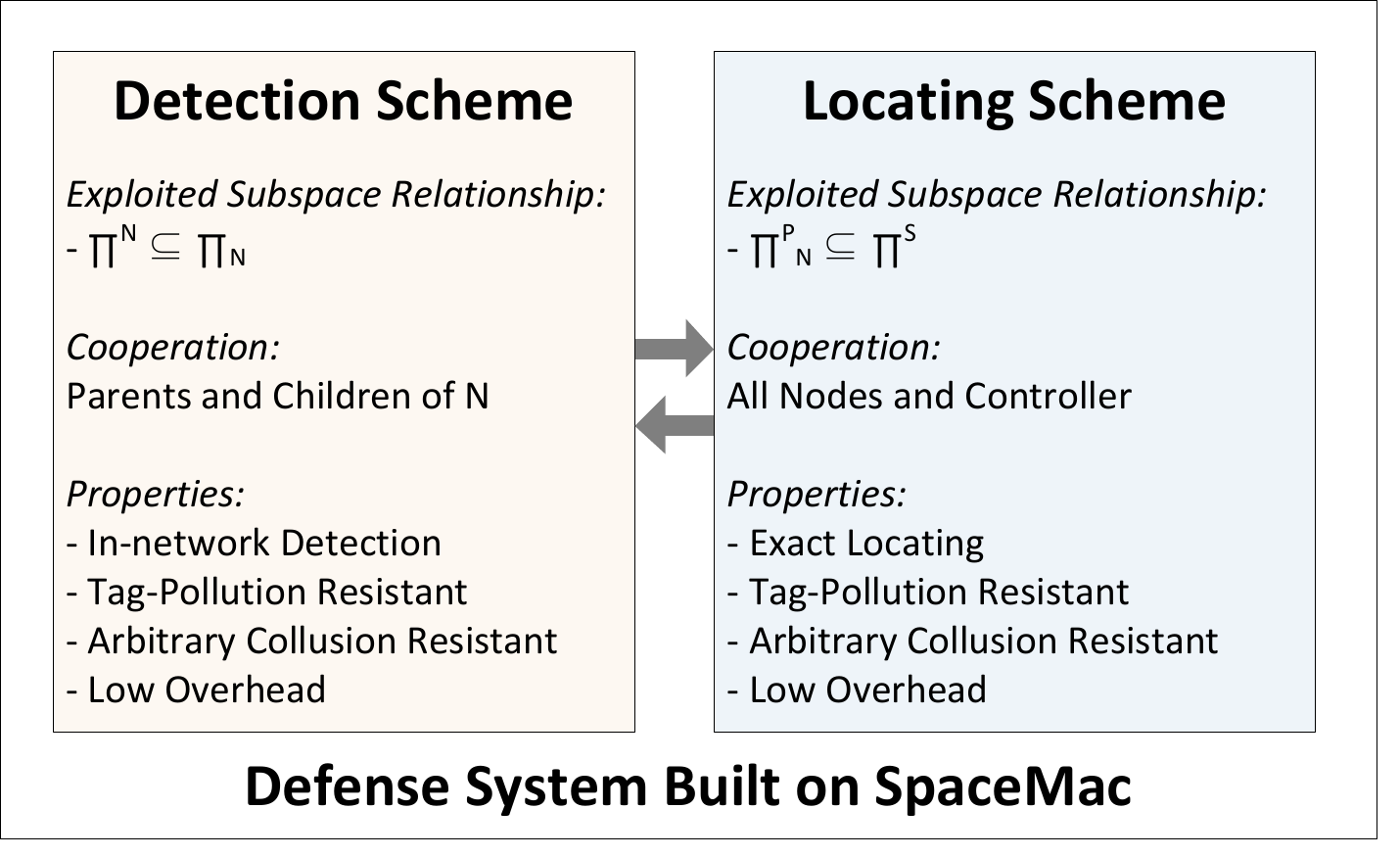}
\vspace*{-5pt}
\caption{Cooperative defense system's main components, their exploited subspace relationships, and their properties}
\label{fig:system}
\afterSpace
\end{figure}

\subsection{Design Goals}
With this threat model in mind, we set out to design a defense system with the following goals:
\begin{enumerate}[(1)]
\item {\em In-network Detection.} Any intermediate node in the network should be able to detect the attack as soon as its malicious parent injects a corrupted packet into the network. This prevents corrupted packets from polluting the downstream edges.
\item {\em Exact Locating.} The location of all pollution attackers should be precisely identified. This allows for the removal of the attackers from the network.
\item {\em Arbitrary Collusion Resistance.} The system should able to cope with multiple pollution attackers when they attack independently  as well as when they collude. In particular, the defense system should be able to remove the attackers from the network even when they collude.
\item {\em Low Overhead.} The defense system should have low computation and low communication overhead. In particular, the  system should require a little amount of computing from the nodes in the network and should not introduce a large amount of traffic, \eg, bandwidth of the MAC tags, to the network.
\end{enumerate}

To achieve the above goals, we design a defense system which consists of two main components: the detection scheme and the locating scheme. The detection scheme provides in-network detection while the locating scheme provides exact locating. Both the detection scheme and locating scheme impose little computing overhead as well as communication overhead. The defense system as a whole is arbitrarily collusion resistant. Fig. \ref{fig:system} illustrates the overall structure of our defense system.

\section{Key Observations and Approach Overview}
\label{sec:observation}

\begin{figure}[t!]
\beforeSpace
\centering
\includegraphics[width=4.5cm]{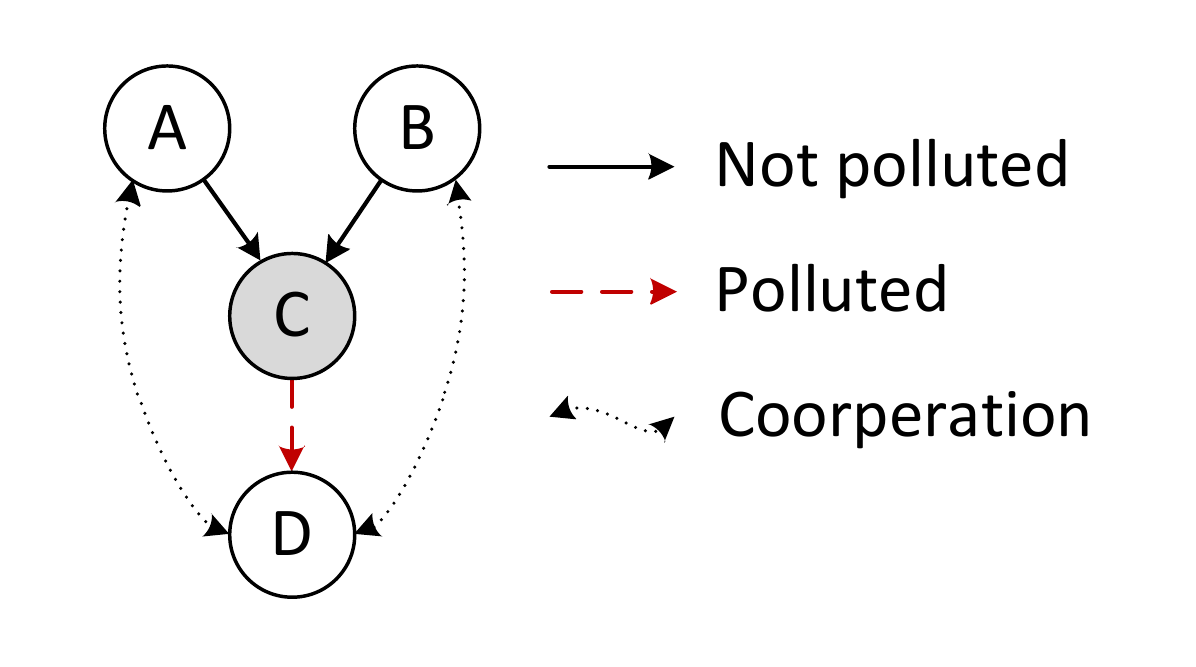}
\vspace*{-15pt}
\caption{An example illustrates how $\SMac$ helps to detect pollution attacks. Using $\SMac$, $A$ and $B$ are able to sign the expanding space $\Pi_C$ (the received space of $C$) and $D$ is able to verify any packet sent by $C$ to see if it belongs to $\Pi_C$. If there is a packet sent by $C$ that is not in $\Pi_C$, the attack is detected by $D$. The cooperation among $A$, $B$, and $D$ helps to detect the attack from $C$.}
\label{fig:detection}
\afterSpace
\end{figure}

\begin{figure}[t!]
\beforeSpace
\vspace*{20pt}
\centering
\includegraphics[width=8cm]{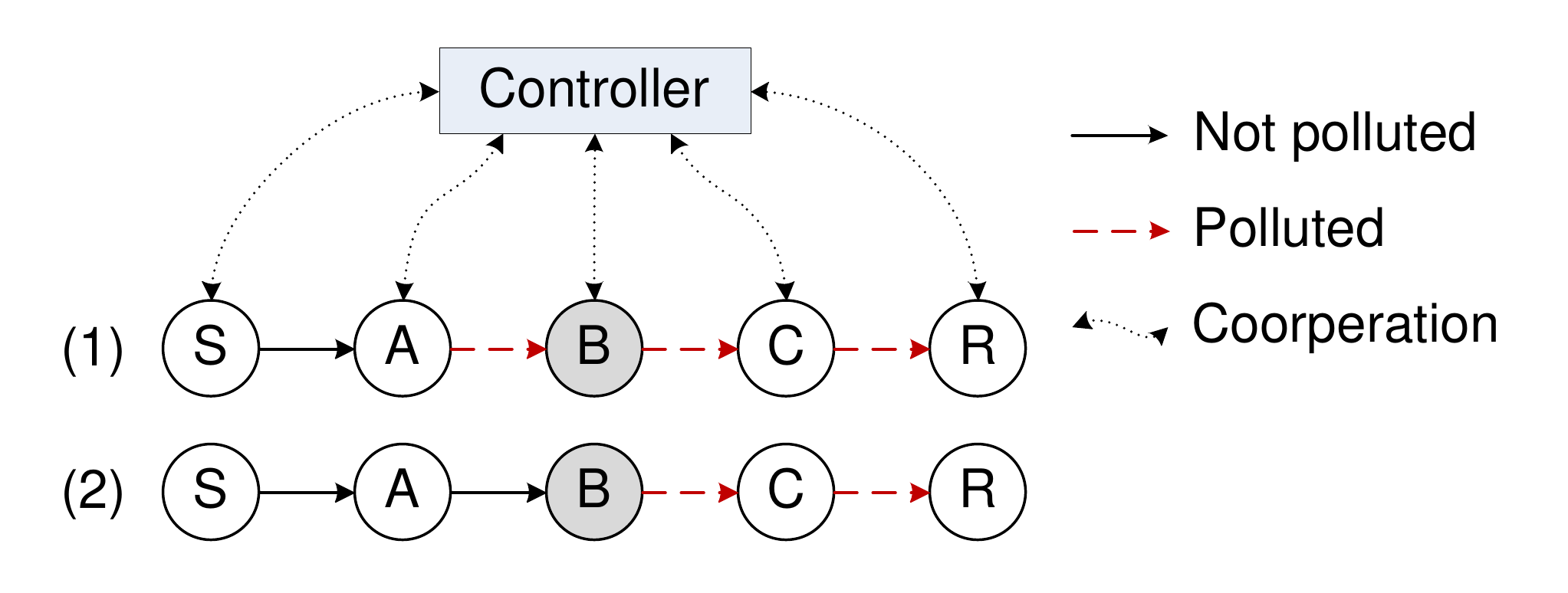}
\vspace*{-15pt}
\caption{An example of inferring an attacker's location using information about polluted edges from \cite{Jafarisiavoshani2008}: The attacker is at node $B$. Scenarios (1) and (2) correspond to the sets of polluted edges when the attacker lies and is honest about its incoming subspace, respectively. The controller can narrow down the attacker to two nodes: $A$ and $B$, as they initiate polluted edges. The cooperation among nodes in the network and the controller helps the locating process.}
\label{fig:locate}
\afterSpace
\end{figure}

In this section, we describe our key observations and how we exploit these observations in our design of the detection and locating schemes.

\subsection{In-Network Detection}
\label{subsec:detection_obs}

Previous work that uses homomorphic MACs to detect corrupted packets, such as, the work by
Agrawal and Boneh \cite{Agrawal2009}, Li \ea~\cite{Li2010}, and Zhang \ea~\cite{Zhang2011}, leverage the observation that if a packet does not belong to the source space, then it is a corrupted packet. The detection works by first establishing shared secret MAC keys between the source and the intermediate nodes. Then, using these secret keys, the source node can sign the fixed source space and the intermediate nodes can verify if their received packets belong to the source space.

Our detection scheme leverages a different observation: A packet sent by an intermediate node must belong to the space spanned by all packets that it received from its parents (Lemma \ref{lem1}). For example, consider a subset of nodes in a network in Fig. \ref{fig:detection}. At any moment in the multicast session, a packet sent by $C$ must belong to the space spanned by the packets it received from its parents: $A$ and $B$; otherwise, $C$ must be polluting the network. Formally, at any moment $t$ in the multicast session, if an intermediate node $N$ sends out a vector $\vct{y}$ then $\vct{y} \in \Pi_N(t)$; otherwise, $\vct{y}$ is corrupted.

We use $\SMac$ to enable the parents of $N$ to sign the expanding space $\Pi_N$ and the children of $N$ to verify any packet that $N$ sends to see if it belongs to $\Pi_N$. The cooperation of the parents of $N$ and the children of $N$ enables the children to detect any corrupted packet sent by $N$ immediately. For example, in Fig. \ref{fig:detection}, at any time $t$, $D$ is able to check if any packet it receives from $C$ belongs to $\Pi_C(t)$; hence, $D$ is able to detect any pollution attempt by $C$ as soon as $D$ receives a corrupted packet from $C$.

\subsection{Exact Locating}
\label{subsec:locate_obs}

Leveraging the cooperation between nodes in the network and a central controller, Jafarisiavoshani \ea~ \cite{Jafarisiavoshani2008} have shown that when there is a single pollution attacker, its location can be narrowed down to a set of at most two nodes. This is by analyzing the polluted edges identified based on the incoming subspaces reported by all the nodes to the central controller. An example is shown in Fig. \ref{fig:locate}. When there are multiple attackers in a general network topology, the number of suspected nodes increases to include the attackers and their parents and children.

Our key observation here is that the uncertainty about the location of the attackers originates from the fact that the attackers can lie about their received spaces. Therefore, by ensuring that all nodes in the network cannot lie about their received spaces, we can exactly locate the attackers. For instance, if the attacker cannot lie in the example given in Fig. \ref{fig:locate}, then the only possible scenario is scenario (2). Thus, one can determine that the attacker is at node B. We explicitly design and use $\SMac$ to achieve this goal, {\em i.e.,} prevent nodes from lying. To prevent attacker $B$ from lying, $A$ cooperates with the controller by signing the space spanned by the packets it sends to $B$ using $\SMac$. This is so that when $B$ reports a fake space, it will not have the proper signature of the fake space to convince the controller.


\section{The Construction of SpaceMac}
\label{sec:construction}
In this section, we describe the construction of $\SMac$. This construction is new and significantly more efficient than our old construction presented in \cite{Le2010}. This construction is an improvement of the homomorphic MAC construction, $\mathsf{HomMac}$, proposed by Agrawal and Boneh \cite{Agrawal2009}. 
The key difference between $\SMac$ and $\mathsf{HomMac}$ is the security property that $\SMac$ brings: {\em $\SMac$ allows for signing spaces that expand over time, while $\mathsf{HomMac}$ only allows for signing fixed spaces}. This is directly reflected in the difference between the security games of $\SMac$ and $\mathsf{HomMac}$, and consequently the difference between the constructions. $\SMac$ improves and generalizes of $\mathsf{HomMac}$. Therefore,  $\SMac$ can also be used to sign fixed spaces as well, \eg,  in the place of $\mathsf{HomMac}$ in the detection scheme in \cite{Agrawal2009}. However,  it is not possible to use $\mathsf{HomMac}$ to support either our detection scheme or our locating scheme because any intermediate node does not know about the entire source space (to use $\sign$ of 
$\mathsf{HomMac}$) at the time of tag generation.


\subsection{Definitions}
A ($q, n, m$) homomorphic MAC scheme is defined by three probabilistic, polynomial-time algorithms: $\mac$, $\cbn$, and $\vrf$. The $\mac$ algorithm generates a tag for a given vector; the $\cbn$ algorithm computes a tag for a linear combination of some given vectors; and the $\vrf$ algorithm verifies whether a tag is a valid tag of a given vector.
\begin{itemize}
\item $\mac(k, \text{id}, \mathbf{y})$:
  \begin{itemize}
  \item Input: a secret key $k$, the identifier $\text{id}$ of the source space $\Pi^S$, and a vector $\mathbf{y} \in \mathbb{F}^{n+m}_q$.
  \item Output: tag $t$ for $\mathbf{y}$.
  \end{itemize}
\item $\cbn((\mathbf{y}_1,t_1,\alpha_1), \cdots, (\mathbf{y}_p,t_p,\alpha_p))$:
  \begin{itemize}
  \item Input: $p$ vectors $\vct{y}_1, \cdots, \vct{y}_p$, their tags $t_1, \cdots, t_p$ under key $k$, and their coefficients $\alpha_1, \cdots, \alpha_p \in \mathbb{F}_q$.
  \item Output: tag $t$ for vector $\vct{y} \overset{\text{def}}{=} \sum_{i=1}^p \alpha_i\,\vct{y}_i$.
  \end{itemize}
\item $\vrf(k, \text{id}, \vct{y}, t)$:
  \begin{itemize}
  \item Input: a secret key $k$,  the identifier $\text{id}$ of the source space $\Pi^S$,  a vector $\mathbf{y} \in \mathbb{F}^{n+m}_q$, and its tag $t$
  \item Output: 0 (reject) or 1 (accept)
  \end{itemize}
\end{itemize}
Also, the scheme must satisfy the following correctness requirement:
{\footnotesize \[\vrf \left( k, \text{id}, \sum_{i=1}^p \alpha_i \vct{y}_i, \cbn((\mathbf{y}_1,t_1,\alpha_1), \cdots, (\mathbf{y}_p,t_p,\alpha_p)) \right) = 1.\]}

\subsection{Attack Game}

We consider the following attack game for a homomorphic MAC $\mathcal{T} = (\mathsf{Mac, Combine, Verify})$, a challenger $\mathcal{C}$, and an adversary $\mathcal{A}$:
\begin{itemize}
\item \emph{Setup.} $\mathcal{C}$ generates a random key $k \overset{\text{R}}{\leftarrow} \mathcal{K}$
\item \emph{Queries.} $\mathcal{A}$ adaptively queries $\mathcal{C}$, where each query is of the form $(\text{id}, \vct{y})$. For each query,  $\mathcal{C}$ replies to $\mathcal{A}$ with the corresponding tag $t \leftarrow \mac(k, \text{id}, \vct{y})$.
\item \emph{Output.}  $\mathcal{A}$ eventually outputs a tuple $(\text{id}_*, \vct{y}_*, t_*)$.
\end{itemize}
Up to the time $\mathcal{A}$ outputs, it has queried $\mathcal{C}$ multiple times. Let $l$ denote the number of times $\mathcal{A}$ queried $\mathcal{C}$ using $\text{id}_*$ and get tags for $l$ vectors, $\vct{y}_{*1}, \cdots, \vct{y}_{*l}$, of these queries. Let $\vct{y}_* = (y_*^{(1)}, \cdots, y_*^{(n+m)})$. We consider that the adversary wins the security game if
\begin{enumerate}[(i)]
\item $(y_*^{(n+1)}, \cdots, y_*^{(n+m)}) \neq \vct{0}$ (trivial forge otherwise),
\item  $\vrf (k, \text{id}_*, \vct{y}_*, t_*) = 1\,,\text{ and}$
\item $\vct{y}_* \notin \mathsf{span}(\vct{y}_{*1}, \cdots, \vct{y}_{*l})\,.$
\end{enumerate}

Let Adv[$\mathcal{A}, \mathcal{T}$] denote the probability that $\mathcal{A}$ wins the above attack game. We define a secure homomorphic MAC scheme as follows:

\begin{defn}
A (q, n, m) homomorphic MAC scheme $\mathcal{T}$ is secure if for all probabilistic polynomial-time adversaries $\mathcal{A}$, \emph{Adv[$\mathcal{A}, \mathcal{T}$]} is negligible.
\end{defn}

\subsection{Construction}
\label{subsec:construction}

Let $\mathcal{K}$ and $\mathcal{D}$ denote the domains of the keys and the id's of the spaces sent by the source, respectively. Let $[n]$ denote $\{1, \cdots, n\}$. We use a pseudorandom generator (PRG) $G$: $\mathcal{K}_G \rightarrow \mathbb{F}^{n+m}_q$ and a PRF $F$: $\mathcal{K}_F \times \mathcal{D} \times [m] \rightarrow \mathbb{F}_q$. A key $k$ for this construction consists of a pair $(k_1, k_2)$, where $k_1 \in \mathcal{K}_G$ and $k_2 \in \mathcal{K}_F$.

\begin{itemize}
\item $\mac(k, \text{id}, \vct{y})$: A tag for a vector {\bf y} = $(y^{(1)}, \cdots, y^{(n+m)})$ using key $k = (k_1, k_2)$ is generated as follows:
   \begin{itemize}
   \item $\vct{r} \leftarrow G(k_1) \in \mathbb F_q^{n+m}$
   \item $b \leftarrow \sum_{j=1}^m \left[ y^{(n+j)} \cdot F(k_2, \text{id},j) \right] \in \mathbb F_q$
   \item $t \leftarrow (\vct{r} \cdot \vct{y}) + b \in \mathbb F_q$
   \end{itemize}
\item $\cbn((\mathbf{y}_1,t_1,\alpha_1), \cdots, (\mathbf{y}_p,t_p,\alpha_p))$: The tag $t$ of $\vct{y}$ is computed as follows:
   \begin{itemize}
   \item $t \leftarrow \sum_{i=1}^p \alpha_i\,t_i \in \mathbb{F}_q$
   \end{itemize}
\item $\vrf(k, \text{id}, \vct{y}, t)$: To verify if $t$ is the valid tag of $\vct{y}$ using key $k=(k_1,k_2)$, we proceed as follows:
   \begin{itemize}
   \item $\vct{r} \leftarrow G(k_1) \in \mathbb F_q^{n+m}$
   \item $b \leftarrow \sum_{j=1}^m \left[ y^{(n+j)} \cdot F(k_2, \text{id},j) \right] \in \mathbb F_q$
   \item $a \leftarrow \vct{r} \cdot \vct{y} \in \mathbb F_q$
   \item If $a+b=t$ output 1; otherwise, output 0
   \end{itemize}
\end{itemize}

The correctness of the scheme is proved as follows. Suppose
\[\vct{y} = (y^{(1)}, \cdots, y^{(n+m)}) = \sum_{i=1}^p \alpha_i \vct{y}_i\,.\]
Then,
\begin{align*}
a+b = \vct{r} \cdot \sum_{i=1}^p \alpha_i \vct{y}_i + \sum_{j=1}^m \left[  \left( \sum_{i=1}^p \alpha_i y_{i}^{(n+j)} \right) \cdot F(k_2, \text{id}, j) \right]\\
= \sum_{i=1}^p \alpha_i \left[ (  \vct{r} \cdot \vct{y}_i ) +  \sum_{j=1}^m \left( y_i^{(n+j)} \cdot F(k_2, \text{id},j) \right) \right] = \sum_{i=1}^p \alpha_i t_i\,.
\end{align*}

Compared to our old $\SMac$ construction \cite{Le2010}, this construction is significantly more efficient. In particular, compared to the old $\mac$ and $\vrf$ algorithms, the new ones use one additional PRG call but significantly less number of PRF calls: $m$ as opposed to $n+m$. Considering that a PRG computation is more efficient than a PRF computation and that in practice,
$n$ is typically an order of magnitude larger than $m$, \eg, $n=2048, m=128$ in a live video streaming system \cite{Wang2007a}, this new construction is one order of magnitude more computationally efficient. 

Compared to $\mathsf{HomMac}$ \cite{Agrawal2009}, our construction replaces the $\mathsf{Sign}$ algorithm, which generates tags for all basis vectors of a fixed space, with the $\mac$ algorithm, which generates a tag for any vector in $\mathbb{F}^{n+m}_q$. The $\cbn$ algorithms show that tags generated by our $\mac$ algorithms can be combined to produce a valid tag for an arbitrary linear combination. We note that both our constructions can be considered as a generalization of the scheme in \cite{Agrawal2009} because the $\mac$ algorithm can substitute  the $\sign$ algorithm, \ie, it can generate valid, combinable tags for all basis vectors. 


\subsection{Security}
The security of $\SMac$ is proven by assuming $F$ is a secure PRF and $G$ is a secure PRG. Let $\mathcal{B}_1$ and $\mathcal{B}_2$ denote a PRF adversary and a PRG adversary, respectively. Let PRF-Adv[$\mathcal{B}_1, F$] and PRG-Adv[$\mathcal{B}_2, G$] denote the advantages in winning the PRF and PRG security games, respectively.\footnote{The definition of PRF and PRG security games can be found in \cite{Katz2007}.}

\begin{thm}\label{thm:SpaceMac}
For any fixed q, n, m, $\SMac$ is a secure (q, n, m) homomorphic MAC assuming F is a secure PRF and G is a secure PRG. In particular, for every homomorphic MAC adversary $\mathcal{A}$, there is a PRF adversary $\mathcal{B}_1$ and a PRG adversary $\mathcal{B}_2$, who have similar running time to $\mathcal{A}$, such that
\emph{\[\text{Adv}[\mathcal{A},\SMac] \leq \text{PRF-Adv}[\mathcal{B}_1, F] + \text{PRG-Adv}[\mathcal{B}_2, G] + \frac{1}{q}\,.\]}
\end{thm}

\begin{IEEEproof}
The proof is by using a sequence of games denoted as Game 0, 1, and 2. Let $W_0$, $W_1$, and $W_2$ denote the events that $\mathcal{A}$ wins the homomorphic MAC security in Game 0, 1 and 2, respectively. Let Game 0 be identical to the Attack Game. Hence,
\begin{equation}\label{eq:w0}
\text{Pr}[W_0] = \text{Adv}[\mathcal{A}, \SMac]\,.
\end{equation}

In Game 1, the PRG $G$ is replaced by a truly random string, {\em i.e.}, to respond to the $\mac$ query, the challenger computes $\vct{r} \overset{\text{R}}{\leftarrow} \mathbb{F}_q^{n+m}$ instead of $\vct{r} \leftarrow G(k_1)$. Everything else remains the same. Then, there exists a PRG adversary $\mathcal{B}_2$ such that
\begin{equation}\label{eq:w0-w1}
|\text{Pr}[W_0] - \text{Pr}[W_1]| = \text{PRG-Adv}[\mathcal{B}_2, G]\,.
\end{equation}

In Game 2, the PRF $F$ is replaced by a truly random function, \emph{i.e.}, to respond to the $\mac$ query, the challenger computes $b \leftarrow \sum_{j=1}^m \left[ y^{(n+j)} \cdot s^{(j)} \right]$, where $s^{(j)} \overset{\text{R}}{\leftarrow} \mathbb{F}_q$ instead of $s^{(j)} \leftarrow F(k_2, \text{id}, j)$. Everything else remains the same. Then, there exists a PRF adversary $\mathcal{B}_1$ such that
\begin{equation}\label{eq:w1-w2}
|\text{Pr}[W_1] - \text{Pr}[W_2]| = \text{PRF-Adv}[\mathcal{B}_1, F]\,.
\end{equation}

The complete challenger in Game 2 works as follows:
{\flushleft \emph{Initialization.}} $\vct{r} \overset{\text{R}}{\leftarrow} \mathbb{F}_q^{n+m}$

{\flushleft \emph{Queries.}} $\mathcal{A}$ adaptively queries $\mathcal{C}$, where each query is of the form $(\text{id}, \vct{y})$. $\mathcal{C}$ replies to query $i$ of $\mathcal{A}$ as follows:\\
\hspace*{0.3 cm} if id is never used in any of the previous queries:\\
\hspace*{0.6 cm} $s_i^{(j)} \overset{\text{R}}{\leftarrow} \mathbb{F}_q \text{ for } j = 1, \cdots, m$\\
\hspace*{0.3 cm} else:\\
\hspace*{0.6 cm} $s_i^{(j)}$'s := the ones used in the previous response\\
\hspace*{0.3 cm} send $t := (\vct{r} \cdot \vct{y}) +  \sum_{j=1}^m \left[ y^{(n+j)} \cdot s^{(j)} \right]$

{\flushleft \emph{Output.}}  $\mathcal{A}$ eventually outputs a tuple $(\text{id}_*, \vct{y}_*, t_*)$. To determine if $\mathcal{A}$ wins the game we compute\\
\hspace*{0.3 cm} if $\text{id}_* = \text{id}_i$ (for some $i$) then \hspace*{0.5 cm} // case (i)\\
\hspace*{0.6 cm} set $s_*^{(j)} := s_i^{(j)} \text{ for } j=1, \cdots, m$\\
\hspace*{0.3 cm} else \hspace*{4.1 cm} // case (ii)\\
\hspace*{0.6 cm} set $s_*^{(j)} \overset{\text{R}}{\leftarrow} \mathbb{F}_q \text{ for } j=1, \cdots, m$\\
Let $l$ denote the number of times $\mathcal{A}$ queried $\mathcal{C}$ using $\text{id}_*$ and get tags for $l$ vectors, $\vct{y}_{*1}, \cdots, \vct{y}_{*l}$, of these queries. We consider that the adversary wins the game, \emph{i.e.}, event $W_2$ happens, if
\begin{align}
~&(y_*^{(n+1)}, \cdots, y_*^{(n+m)}) \neq \vct{0}\,,\\
~&t_* = (\vct{r} \cdot \vct{y}_*) + \sum_{j=1}^m \left[ y_*^{(n+j)} \cdot s_*^{(j)} \right],\text{ and}\label{eq:t}\\
~&\vct{y}_* \notin \mathsf{span}(\vct{y}_{*1}, \cdots, \vct{y}_{*l})\label{eq:y}\,.
\end{align}

In the following steps, we show that Pr[$W_2$] = $\frac{1}{q}$. Let $T$ be the event that $\mathcal{A}$ outputs the tuple with a completely new $\text{id}_*$, \emph{i.e.}, $\mathcal{A}$ never made queries using $\text{id}_*$ before.
\begin{itemize}
\item When T happens, \emph{i.e.}, in case (ii), since $s_*^{(j)}$'s are indistinguishable from random values in $\mathbb{F}_q$, and $y_*^{(n+j)}$'s are not all zeros, the right hand side of equation (\ref{eq:t}) is a completely random value in $\mathbb{F}_q$, independent of the adversary's view. Thus,
\begin{equation}\label{eq:w2}
\text{Pr}[W_2 \wedge T] = \frac{1}{q}\,\text{Pr}[T]\,.
\end{equation}

\item When T does not happen, \emph{i.e.}, in case (i), $s_*^{(j)}$ of equation (\ref{eq:t}) have been used to generate tags for vectors $\vct{y}_{*1}, \cdots, \vct{y}_{*l}$.
In this case, we will proceed by showing that when $\vct{y}_* \notin \mathsf{span}(\vct{y}_{*1}, \cdots, \vct{y}_{*l})$, the right hand side of equation (\ref{eq:t}) looks indistinguishable from a random value in $\mathbb{F}_q$. To this end, let $\Pi = \mathsf{span}(\vct{y}_{*1}, \cdots, \vct{y}_{*l})$, and $d$ be the dimension of $\Pi$. Note that $d < n+m$ because otherwise $\Pi = \mathbb{F}_q^{n+m}$, which implies $\vct{y}_* \in \Pi$. Let $\{ \vct{b}_1, \cdots, \vct{b}_d \}$ be a basis of $\Pi$.
Denote $\mathsf{aug}({\vct y})$ as the augmentation of vector {\bf y}, {\em i.e.}, 
$\mathsf{aug}(\vct{y}) = (y^{(n+1)}, \cdots, y^{(n+m)})\,.$

\hspace*{0.3 cm}
{\bf Case (a)}: Consider the case when $\mathsf{aug}(\vct{y}_*)$ can be expressed as a linear combination of $\mathsf{aug}(\vct{b}_i), i \in [1,d]$. Let $\mathsf{aug}(\vct{y}_*) = \sum_{i=1}^d \alpha_i\,\mathsf{aug}(\vct{b}_i)$ for some $\alpha_i$. If we let $\vct{y}' = \sum_{i=1}^d \alpha_i \, \vct{b}_i$, then the valid tag of $\vct{y}$' for the same space $\text{id}_*$ as $\vct{y}_*$ is
\begin{align}
t' &= (\vct{r} \cdot \vct{y}') + \sum_{j=1}^m \left[ (\sum_{i=1}^d \alpha_i \, \vct{b}_i)^{(n+j)} \cdot s_*^{(j)} \right] \nonumber\\
~ &= (\vct{r} \cdot \vct{y}') + \sum_{j=1}^m \left[ \vct{y}_*^{(n+j)} \cdot r_*^{(j)} \right] \label{eq:t'}
\end{align}
By subtracting equation (\ref{eq:t'}) from (\ref{eq:t}), we know that by producing a valid forgery, the adversary found a $\vct{y}_*$ and $t_*$ that satisfy the following equation:
\begin{align}
t_* - t' = \vct{r} \cdot ( \vct{y}_* - \vct{y}') \label{eq:t-t'}
\end{align}
However, since $\vct{y}_* \neq \vct{y}'$ ($\vct{y}'$ is in $\Pi$ but $\vct{y}_*$ is not), and $\vct{r}$ is indistinguishable from a random vector in $\mathbb{F}_q^{n+m}$, the probability that he can satisfy (\ref{eq:t-t'}) is exactly $\frac{1}{q}$.

\hspace*{0.3 cm}
{\bf Case (b)}: Consider the case when $\mathsf{aug}(\vct{y}_*)$ cannot be expressed as a linear combination of  $\mathsf{aug}(\vct{b}_i)$'s. In this case, we proceed by showing that given a fixed $\vct{y}_*$, from the perspective of the adversary, the valid tag $t_*$ of $\vct{y}_*$ is indistinguishable from a random value in $\mathbb{F}_q$:

\hspace*{0.3 cm} Let $s_*^{(i)}, i\in[1,m]$, be the unknowns, and $\vct{s} = (s_*^{(1)}, \cdots, s_*^{(m)})$. By the previous $l$ queries, the adversary learns the following system of $l$ equations and $m$ unknowns:
\begin{align*}
\text{(I)}
\begin{cases}
\mathsf{aug}(\vct{y}_{*1}) \cdot \vct{s} =  t_{\vct{y}_{*1} } - \vct{r} \cdot \vct{y}_{*1}\\
\cdots\\
\mathsf{aug}(\vct{y}_{*l}) \cdot \vct{s} =  t_{\vct{y}_{*l} } - \vct{r} \cdot \vct{y}_{*l}
\end{cases}
\end{align*}

\hspace*{0.3 cm} Since $\{ \vct{b}_1, \cdots, \vct{b}_d \}$ is a basis of $\Pi=\mathsf{span}(\vct{y}_{*1}, \cdots, \vct{y}_{*l})$, the above system is equivalent to the following system of $d$ equations:
\begin{align*}
\text{(II)}
\begin{cases}
\mathsf{aug}(\vct{b}_1) \cdot \vct{s} = u_1\\
\cdots\\
\mathsf{aug}(\vct{b}_d) \cdot \vct{s} = u_d
\end{cases}
\end{align*}
where each $u_j$ is a linear combination of right-hand-side values of the equations of system (I).

\hspace*{0.3 cm} Let $\Pi' = \mathsf{span}(\mathsf{aug}(\vct{b}_1), \cdots, \mathsf{aug}(\vct{b}_d) )$ and $d'$ be the dimension of $\Pi'$. Note that $d' \leq d$. Let $\{ \vct{c}_1, \cdots, \vct{c}_{d'} \}$ be a basis of $\Pi'$. Note that since $\mathsf{aug}({\vct{y}_*}) \notin \Pi'$, $\mathsf{aug}({\vct{y}_*})$ cannot be expressed as a linear combination of $\vct{c}_i$'s. The system of equations (II) is equivalent to the following system of $d'$ equations:
\begin{align*}
\text{(III)}
\begin{cases}
\vct{c}_1 \cdot \vct{s} = e_1\\
\cdots\\
\vct{c}_{d'} \cdot \vct{s} = e_{d'}
\end{cases}
\end{align*}
where each $e_j$ is a linear combination of right-hand-side values of the equations of system (II). A valid tag $t_*$ of $\vct{y}_*$ satisfies the following equation:
\begin{align}
\mathsf{aug}(\vct{y_*}) \cdot \vct{s} = t_* - \vct{r} \cdot \vct{y}_* \,. \label{eq:t*}
\end{align}

\hspace*{0.3 cm} Without loss of generality, assume that the adversary knows $\vct{r}$. Note that $d < m$ otherwise $\mathsf{aug}(\vct{y}_*) \in \mathsf{span}(\vct{b}_i)$. Since $d' \leq d$, it follows that $d' < m$. The system of $m$ unknowns and $d'+1$ linear equations, $d'$ from (III) and 1 from (\ref{eq:t*}), is consistent regardless of the value of $t_*$ because the coefficient matrix, whose rows are linearly independent vectors: $\vct{c_1}, \cdots, \vct{c}_{d'}$, and $\mathsf{aug}(\vct{y}_*)$, has rank $d'+1 \leq m$. Furthermore, for any value $t_*$, the the solution space always has size $q^{m-d'-1}$. Thus, for a fixed $\vct{y}_*$, its valid tag, $t_*$, could be any value in $\mathbb{F}_q$ equally likely, given that $s_*^{(i)}$'s are chosen uniformly at random from $\mathbb{F}_q$. Hence, the probability of forging a valid tag $t_*$ is $\frac{1}{q}$.

\hspace*{0.3 cm} By the result of case (a) and case (b),
\begin{equation}\label{eq:w2'}
\text{Pr}[W_2 \wedge \neg T] = \frac{1}{q}\,.\,\text{Pr}[\neg T]\,.
\end{equation}

\item From equations (\ref{eq:w2}) and (\ref{eq:w2'}), we have
\begin{equation}\label{eq:w2f}
\begin{split}
\text{Pr}[W_2] &= \text{Pr}[W_2 \wedge T] + \text{Pr}[W_2 \wedge \neg T] = \frac{1}{q}\,.
\end{split}
\end{equation}
\end{itemize}

Equations  (\ref{eq:w0}), (\ref{eq:w0-w1}), (\ref{eq:w1-w2}), and (\ref{eq:w2f}) together prove the theorem.
\end{IEEEproof}

Theorem \ref{thm:SpaceMac} states that an adversary $\mathcal{A}$ can break the scheme with probability $\frac{1}{q}$. For a small field size, \eg, $q=2^8$, the security of the MAC scheme may be unsatisfactory. To improve the security, one could either increase the field size or use multiple tags as suggested in \cite{Agrawal2009} and \cite{Li2010}. The security of our scheme using $l$ tags is $(\frac{1}{q})^l$. As observed in \cite{Li2010}, it is preferable to use multiple tags instead of increasing the field size. This is because in order to achieve the same security $(\frac{1}{q})^l$, using the field size ${q^l}$ instead of using $l$ tags increases the computational complexity of field multiplication by $\text{log}\,l$ times.

\section{Detection Scheme}
\label{sec:detection}

\begin{figure}[t!]
\beforeSpace
\centering
\includegraphics[width=4cm]{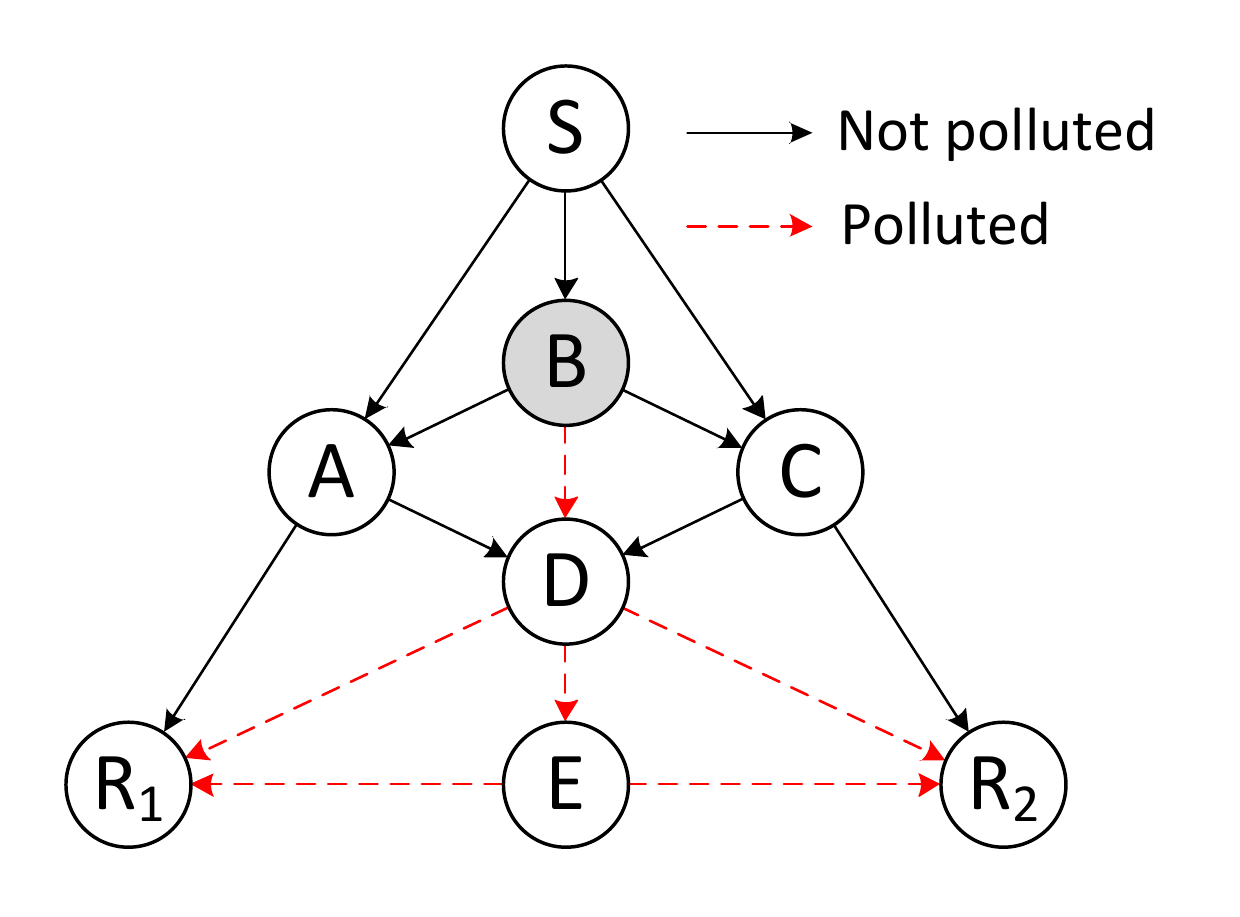}
\vspace*{-10pt}
\caption{A network consisting of 8 nodes (resembling a network given in \cite{Jafarisiavoshani2008}). $B$ is the attacker. After every node (except for $S$) reports to the controller its true incoming spaces, $B$ is identified as the attacker since it has no incoming polluted edge but has one outgoing polluted edge.}
\label{fig:ex_network}
\afterSpace
\end{figure}

In this section, we describe our detection scheme in detail. Our scheme exploits the observation outlined in Section \ref{subsec:detection_obs} to provide in-network detection. In particular, parents and children of an intermediate node $N$ cooperate through $\SMac$ to detect any corrupted packet sent by $N$ in a timely manner. For ease of presentation, we describe the detection scheme within the scope of a single generation, \ie, considering a single source space {id}.

{\bf 1) Assumptions.}\quad
We assume that there is a controller who knows the complete topology of the graph. The controller could be the source itself. This assumption is also made in recent work by Li \ea~\cite{Li2010}. We further assume that each node $N$ shares with the controller a pair of secret keys $(k^1_N, k^2_N)$. These keys can be established with the help of a Public Key Infrastructure (PKI). We note that recent proposed schemes which use homormorphic MACs also made assumption about the existence of a PKI \cite{Li2010, Zhang2011} or the existence of shared secret keys \cite{Agrawal2009}. In general, the problem of establishing shared secret keys is a challenging problem of its own and is orthogonal to this work.

{\bf 2) Bootstrapping.}\quad
First, for every intermediate node $N$, the controller determines the key $k_{\bar{N}}$, which will be secret to $N$ itself and is used by the parents and children of $N$ when using $\SMac$. Each node can serve as either a parent or a child; therefore, each node, depending on its position in the network, requires to know a different set of keys to participate in the detection scheme. For example, consider the network in Fig. \ref{fig:ex_network}, node $D$ needs to know $k_{\bar{A}}, k_{\bar{B}}$, and $k_{\bar{C}}$ to detect corrupted packets sent by $A$, $B$ and $C$, respectively. It also needs to know $k_{\bar{E}}$ to help $R_1$ and $R_2$ to detect corrupted packets sent by $E$. 

Second, the source and all the receivers need to share an end-to-end key, $k^*$. This key is used to ensure detection in the presence of colluding adversaries, in which case a node $N$ colludes with its parent to obtain $k_{\bar{N}}$ and thus can bypass the verification of its children. We defer the discussion about multiple adversaries to Section \ref{subsec:multiple_adversaries}, where we analyze different colluding scenarios in depth.

The controller then sends to each node $N$ a bootstrapping packet consisting of the set of keys that are necessary for it to participate in the detection scheme. In particular, the controller construct the bootstrapping packet $b_N$: 
$b_N = \{ k_{\bar{X}, X \in \{ \mathcal{P}_N \cup \mathcal{C}_N \} }, k^* \}\,.$
Note that $b_N$ contains $k^*$ if and only if $N$ is either the source or a receiver. The controller then sends $b_N$ to $N$ through a secure and authenticated channel achieved with $k^1_N$ (for encryption) and $k^2_N$ (for authentication) using a standard encrypt-then-authenticate algorithm\footnote{We refer the reader to Chapter 4.9 in \cite{Katz2007} for more details on encrypt-then-authenticate algorithms.}. For example, node $D$ in Fig. \ref{fig:ex_network} receives $\{ k_{\bar{A}}, k_{\bar{B}}, k_{\bar{C}}, k_{\bar{E}} \}$ while node $R_1$ receives $\{ k_{\bar{A}}, k_{\bar{D}}, k_{\bar{E}}, k^* \} $. 

We note that the MAC keys $k^*$ and $k_{\bar{X}}$ can be used for multiple source space/generations. This is because, by construction, $\SMac$ takes into account generation identifiers when computing tags. As a result, the overhead of a key establishment is per multiple generations as opposed to per single generation. Hence, this overhead is asymptotically negligible in the number of generations.

Finally, if  each node knows its own 2-hop neighborhood information, then the MAC keys can also be bootstrapped without the help of the controller. In particular, by using a secure key distribution scheme for ad-hoc networks, such as \cite{Khalili2003}, the source can establish the shared secret MAC key $k^*$ with the receivers, and for a node $N$, a parent $P$ can establish a shared secret MAC key $k_{\bar{N}}$ with the parents and children of $N$.

{\bf 3) Sending and Coding.}\quad
Before sending out the each source packet, $\vct{v}_i$, the source $S$ calculates an {\em end-to-end tag}, $t^{k^*}_{\vct{v}_i}$, using the $\mac$ algorithm of $\SMac$ with key $k^*$:
$t^{k^*}_{\vct{v}_i} = \mac(\vct{v}_i, k^*)\,.$ 
$S$ then attaches this tag to every source packet and sends $\vct{w}_i \triangleq \{ t^{k^*}_{\vct{v}_i} \, || \, \vct{v}_i \}$ instead of $\vct{v}_i$,  where `$||$' denotes concatenation. The packets traversing the network are linear combinations of $\vct{w}_i$'s instead of $\vct{v}_i$'s. For ease of presentation with regards to the input length of the $\SMac$ algorithms, assume that the size of $\vct{w}_i$ is still $n$.

Consider a parent $P$ who wants to send a packet $\vct{y}$ to its child $N$. $P$ needs to calculate a {\em helper tag} which helps the children of $N$ to detect corrupted packets sent by $N$. In particular, before sending $\vct{y}$ to $N$, $P$ needs to calculate a MAC tag, $t^{k_{\bar{N}}}_{\vct{y}}$, using $\mac$ under key $k_{\bar{N}}$:
$t^{k_{\bar{N}}}_{\vct{y}} = \mac( \vct{y}, k_{\bar{N}} )\,.$ 
Besides the helper tag, $P$ must also pass along a {\em verification tag} of $\vct{y}$, which is used by $N$ to verify the integrity of $\vct{y}$. Assume $P$ received $\{ \vct{y}_1, \cdots, \vct{y}_l \}$ and their helper tags $\{ t^{k_{\bar{P}}}_{\vct{y}_1}, \cdots, t^{k_{\bar{P}}}_{\vct{y}_l} \}$ from its parents, and $P$ computes $\vct{y}$ as $\vct{y} = \sum_{i=1}^l \alpha_i \vct{y}_i$. Then, the verification tag, $t^{k_{\bar{P}}}_{\vct{y}}$, of $\vct{y}$ can be computed using $\cbn$:
$t^{k_{\bar{P}}}_{\vct{y}} = \cbn( (\vct{y}_1,  t^{k_{\bar{P}}}_{\vct{y}_1}, \alpha_1), \cdots, (\vct{y}_l,  t^{k_{\bar{P}}}_{\vct{y}_l}, \alpha_l) )\,.$
The final packet that $P$ sends to $N$ includes $\vct{y}$ and its helper and verification tags:
$\{ t^{k_{\bar{N}}}_{\vct{y}} \,||\,  t^{k_{\bar{P}}}_{\vct{y}} \,||\, \vct{y} \} \,.$

We note that if a node is benign, besides explicitly calculating the helper tag, it would code and send packets in a way identical to what it does normally. The verification tag will be computed correctly because the $\cbn$ algorithm linearly combines the tags in the same way the packets are combined, \ie, with the same set of coefficients.

{\bf 4) Receiving and Verification.}\quad
When a node $N$ receives from its parent $P$ a packet $\{ t^{k_{\bar{N}}}_{\vct{y}} \,||\,  t^{k_{\bar{P}}}_{\vct{y}} \,||\, \vct{y} \}$, it uses $k_{\bar{P}}$ and the $\vrf$ algorithm to verify the integrity of the packet. The packet is deemed non-corrupted if
$\vrf (k_{\bar{P}}, \vct{y}, t^{k_{\bar{P}}}_{\vct{y}}) = 1\,.$ 
The security guarantee comes from the security of $\SMac$: since $P$ does not know $k_{\bar{P}}$, the probability that $P$ can forge a valid tag, $t^{k_{\bar{P}}}_{\vct{y}}$, when $\vct{y}$ is outside of its received space, $\Pi_P$, is $\frac{1}{q}$. As a result, as soon as $N$ receives a corrupted packet from $P$, with high probability, $N$ is able to detect the attack immediately.

In the case $N$ is a receiver, it further verifies the end-to-end tag using key $k^*$. Parse $\vct{y}$ as $\{ t^{k^*}_{\vct{w}} \,||\, \vct{w}\}$. $N$ accepts $\vct{w}$ if
$\vrf (k^*, \vct{w}, t^{k^*}_{\vct{w}}) = 1\,.$ 
The security guarantee, again, comes from the security of $\SMac$: since none of the malicious intermediate node knows $k^*$, if $\vct{w}$ is outside of the source space, the adversary can only forge a valid tag of $\vct{w}$, $t^{k^*}_{\vct{w}}$, with a negligible probability of $\frac{1}{q}$. This second level of verification is to provide a detection mechanism in the presence of colluding adversaries.

\section{Locating Scheme}
\label{sec:locating}

Locating the attackers to eventually eliminate them is a logical step after a pollution attack is detected. In this section, we describe in detail how we exploit the observation made in Section \ref{subsec:locate_obs} to exactly locate the pollution attackers. The main idea is to force nodes to truthfully report their received spaces to correctly identify polluted edges, thereby enabling the exact identification of the location of the attackers.

\subsection{Overview}
\label{subsec:locating_overview}

{\bf 1) Reporting.}\quad
The following lemma, originally presented in \cite{Jafarisiavoshani2007} and \cite{Jafarisiavoshani2008}, implies that for each received subspace, $\Pi_N^P$, from a parent $P$, node $N$ may report a randomly chosen packet, $\mathbf{y}_r$, of the space instead of the space itself; and by checking if $\mathbf{y}_r \in \Pi^S$, the controller can determine if $\Pi_N^P \subseteq \Pi^S$ to identify the polluted edges.

\begin{lem}[Jafarisiavoshani \emph{et al.} \cite{Jafarisiavoshani2008, Jafarisiavoshani2007}]\label{lem:subspace}
Let $\Pi_1$ and $\Pi_2$ be two subspaces of $\mathbb{F}^{n+m}_q$ and assume that $\mathbf{y}_r$ is a randomly selected packet from $\Pi_1$. Let $d_{12}$ and $d_1$ are the dimensions of $\Pi_1 \cap \Pi_2$ and $\Pi_1$, respectively. With probability $1-q^{d_{12}-d_1}$, $\mathbf{y}_r \in \Pi_2$ if and only if $\Pi_1 \subseteq \Pi_2$.
\end{lem}

{\bf 2) Using $\SMac$.}\quad
We use $\SMac$ to prevent nodes from lying about their received spaces as follows. To enforce a node $N$ to report a true received space,  $\Pi_N^P$, that it received from its parent, $P$, the parent $P$ and the controller cooperate so that the controller only accepts reported packets belonging to but not outside of $\Pi_N^P$. In particular, whenever $P$ sends a vector $\vct{y}_i$ to $N$, it generates a tag, $t_{\vct{y}_i}$, of $\vct{y}_i$ using the $\mac$ algorithm with a secret key shared by $P$ and the controller. Then, when $N$ reports $\mathbf{y}_r$, if $\mathbf{y}_r$ is a linear combination of vectors that it received from $P$, $\vct{y}_i$'s, then $N$ can generate a valid tag for $\mathbf{y}_r$ by using the $\cbn$ algorithm on the tags of $\vct{y}_i$'s that it received; if $\mathbf{y}_r$ is not a linear combination of $\vct{y}_i$'s then $N$ can forge a valid tag for $\vct{y}_r$ with only a negligible probability of $\frac{1}{q}$.

{\bf 3) Non-Repudiation Transmission Protocol.}\quad
As presented, $\mathsf{SpaceMac}$ forces nodes to report only true received subspaces since it is computationally difficult to forge valid tags otherwise. However, it does not prevent a malicious node from sending invalid tags to its children to prevent the children from reporting polluted spaces.

For example, an attacker $P$ can send a polluted packet $\vct{y}_e \notin \Pi^S$ and a bogus tag $t_e$ to its child $N$. When $N$ reports the space $\Pi_N^P$, if the randomly chosen vector $\vct{y}_r$ was formed by a linear combination involving $\vct{y}_e$, then the aggregated tag $t_r$ of $\vct{y}_r$ that $N$ generates using the $\cbn$ algorithm will be invalid due to the bogus tag $t_e$. As a result, the controller will reject $\vct{y}_r$. Consequently, the attacker $P$ successfully prevents its benign child $N$ from reporting the polluted space $\Pi_N^P$.

To address this, we utilize an efficient non-repudiation transmission protocol proposed by Wang \emph{et al.} \cite{Wang2010}. For a parent $P$ and a child $N$, the controller generates a set of secret keys, denoted by $\mathcal{X}$, based on the private key of the parent and the ID $N$ of the child. After that, the controller randomly selects a set of keys $\mathcal{Y}$ from $\mathcal{X}$ based on the private key of the child and the ID $P$ of the parent; then, it sends $\mathcal{Y}$ to the child. We denote $\mathcal{X} \setminus \mathcal{Y}$ as $\overline{\mathcal{Y}}$; also, let $\lambda \triangleq |\mathcal{X}|$ and $\delta \triangleq |\mathcal{Y}|$.

When sending a packet, $P$ generates $\lambda$ tags (instead of one) using the $\mac$ algorithm and all keys in $\mathcal{X}$. When receiving a packet, $N$ uses its set of keys $\mathcal{Y}$ and the $\vrf$ algorithm to verify $\delta$ out of $\lambda$ tags. Finally, when receiving a randomly chosen packet $\vct{y}_r$ chosen from $\Pi_N^P$ and its $\lambda$ tags from the $N$, the controller uses all keys in $\overline{\mathcal{Y}}$ and the $\vrf$ algorithm to verify all $\lambda - \delta$ tags. The controller, in this case, keeps track of a counter $\theta$, $\theta \leq \lambda - \delta$. If at least $\theta$ tags pass the verification then the controller accepts the report. 



The following two lemmas provide the security of the non-repudiation transmission protocol when applying to our context. Lemma \ref{lem:non-reput-rcvr} is identical to Theorem 1 in \cite{Wang2010}. Lemma \ref{lem:non-reput-sndr} is an adapted version of Theorem 2 in \cite{Wang2010} -- the difference is that in our case, a node does not report a packet that it receives, but it reports a linear combination of packets that it receives instead.

\begin{lem}[Non-repudiation of the receiver--Wang \emph{et al.} \cite{Wang2010}]\label{lem:non-reput-rcvr}
The probability that a malicious child node can successfully report to the controller that its parent sends it a packet $\vct{y}$, which is never sent by the parent, is at most
\begin{equation*}
\sum_{i=\theta}^{\lambda - \delta} \binom{\lambda - \delta}{i} \, \frac{1}{q^i} \, \left( 1 - \frac{1}{q} \right) ^ {\lambda - \delta - i}\,.
\end{equation*}
\end{lem}



\begin{lem}[Non-repudiation of the sender--Wang \emph{et al.} \cite{Wang2010}]\label{lem:non-reput-sndr}
The probability that a malicious parent can make the controller reject the parent's space reported by its child by sending the child some packets with invalid tags is at most
\[\underset{0 \leq x \leq \delta+\theta-1}{\emph{max}} \,p(x),\text{ where } p(x) \leq  \sum_{i= \text{max}(x - \theta + 1,0)}^{\text{min}(\delta,x)} \frac{\binom{\delta}{i}  \, \binom{\lambda-\delta}{x-i} }{\binom{\lambda}{x} \, q^{\delta - i}}\,.\]
\end{lem}

The proofs of these two lemmas are provided in the Appendix. Finally, we note that both of the above probabilities can be made very small by choosing appropriate values for  $q$, $\lambda$, $\delta$, and $\theta$. Examples of values for these parameters and the corresponding probabilities are provided in Table \ref{table:params}. The choice of parameters can then be made based on the desired tradeoff between the overhead and the probability that the attacker succeeds.

{\bf 4) Locating the Attackers.}\quad
After the controller collects the true subspaces from every node, we proceed similar to the approach by Jafarisiavoshani \emph{et al.} \cite{Jafarisiavoshani2008} to locate the attackers. Here, we discuss the case when there is a single attacker. We defer the case when there are multiple attackers to Section \ref{subsec:multiple_adversaries}.

In \cite{Jafarisiavoshani2008}, the authors have shown that in a general network which has a single adversary, the location of the adversary can be narrowed down to a set of at most two nodes in both cases where the adversary inject corrupted packets to either one downstream edge or multiple downstream edges. This is done by partitioning the edges into two set: the set of polluted edges, $\mathcal{E}_p$, and non-polluted edges, $\mathcal{E}_s$, then analyzing the nodes with respect to the identified $\mathcal{E}_p$ and $\mathcal{E}_s$. They also note that the partitioning of $\mathcal{E}_p$ and $\mathcal{E}_s$ is not unique since the adversary might lie, which results in the uncertainty about the location of the attacker.

Fortunately, when the partition reflects the real state of pollution of the edges in the network, \emph{i.e.}, when the adversary is forced to report its true incoming spaces, the adversary is always the node that has no incoming edge belonging to $\mathcal{E}_p$ but has at least one outgoing edge belonging to $\mathcal{E}_p$. Fig. \ref{fig:ex_network} shows the case where $B$ is an attacker whom get identified because it has no incoming polluted edge but one outgoing polluted edge.

Using our scheme, the probability that the attacker lies about its incoming spaces is very small (Lemma \ref{lem:non-reput-rcvr}). Furthermore, the probability that the attacker can prevent its children from reporting the subspaces polluted by itself is very small, too (Lemma \ref{lem:non-reput-sndr}). As a result, with high probability (depending on $q$, $\lambda$, $\delta$, and $\theta$), our scheme can produce an unambiguous partitioning of $\mathcal{E}_p$ and $\mathcal{E}_s$, which helps to precisely locate the attacker.

\subsection{Full Description}
To distinguish cryptographic keys used in the detection scheme and keys used in the locating scheme, we decorate any key used in the locating scheme with an overhead bar, \eg, $\bar{k}$.

{\bf 1) Assumptions.}\quad
Similar to the assumptions we made in the detection scheme, we assume that there is a controller (could be the source itself) who knows the complete topology and the source space. This assumption is also made in recently proposed locating schemes \cite{Jafarisiavoshani2008, Wang2010}. We assume that each node $N$ shares a triplet of secret keys $(\bar{k}^1_N, \bar{k}^2_N, \bar{k}^3_N)$ with the controller (with the help of a PKI). We further assume that each node knows the identifiers of its adjacent nodes (can be bootstrapped by the controller). In addition, we assume that there is a reliable low-bandwidth end-to-end communication path between the controller and each node (the channel for the reports and the announcement by the controller). Other locating schemes, such as, \cite{Jafarisiavoshani2008} and \cite{Wang2010}, implicitly made this assumption.

{\bf 2) Bootstrapping.}\quad
Let $\mathcal{N}$ be the set of IDs of adjacent downstream nodes of $P$. For $N \in \mathcal{N}$, the controller generates a set $\mathcal{X}_{PN}$ of $\lambda$ keys using a PRF $F_1$:
$\mathcal{K} \times \mathcal{I} \times [\,\lambda\,] \rightarrow \mathcal{K}$, where $\mathcal{K}$ is the domain of key $\bar{k}^1_P$, and $\mathcal{I}$ is the domain of the identifiers of the nodes:
$\mathcal{X}_{PN} \leftarrow \{F_1 (\bar{k}^1_P, N, i), \text{ for } i = 1, \cdots, \lambda \}\,.$ Note that $P$ can compute $\mathcal{X}_{PN}$ itself as it knows $\bar{k}^1_P$ and its neighbors' identifiers.

For $N \in \mathcal{N}$, consider an array $L$ whose elements are distinct subsets of size $\delta$ of $\mathcal{X}_{PN}$. Note that $L$ has length $\binom{\lambda}{\delta}$. The controller uses another PRF $F_2$: $\mathcal{K} \times \mathcal{I} \rightarrow [\binom{\lambda}{\delta}]$  to select from $L$ a subset of size $\delta$: $\mathcal{Y}_{PN} = L[i], \text{ where } i \leftarrow F_2 (\bar{k}^1_N, P)$. The controller then sends $\mathcal{Y}_{PN}$ to node $N$ through a secure and authenticated channel achieved with $\bar{k}^2_N$ (for encryption) and $\bar{k}^3_N$ (for authentication) using an encrypt-then-authenticate algorithm. Note that similar to $k^*$ and $k_{\bar{X}}$, the sets of keys $\mathcal{X}_{PN}$ and $\mathcal{Y}_{PN}$ can be used across multiple generations. Denote $\mathcal{X}_{PN} \setminus \mathcal{Y}_{PN}$ as $\overline{\mathcal{Y}_{PN}}$.

{\bf 3) Sending and Receiving.}\quad
Let id be the identifier of the current source space $\Pi^S$. When a node $P$ sends a packet $\vct{y}$ to its downstream node $N$, beside the id, it has to send along $\lambda$ tags, which are computed using the $\mac$ algorithm and keys in $\mathcal{X}_{PN}$. Let $\mathcal{G}_{PN}(\vct{y})$ denote this set of tags.
Node $P$ sends $(id, \vct{y}, \mathcal{G}_{PN}(\vct{y}))$. When node $N$ receives this packet from node $P$, it uses $\mathcal{Y}_{PN}$ and the $\vrf$ algorithm to check the validity of $\delta$ out of $\lambda$ tags of $\mathcal{G}_{PN}(\vct{y})$. It drops $\vct{y}$ as long as there is an invalid tag. Otherwise, it stores the received tuple in its buffer.

{\bf 4) Pollution Detection and Alert.}\quad
A detection of the pollution is needed to start the locating process. Here, we use our detection scheme to provide the detection. Nevertheless, we stress that our locating scheme does not depend on any particular detection scheme. Using our detection scheme, a node $N$, upon detecting a pollution, sends an alert (id $||$ $N$) to the controller through an authenticated channel achieved using a traditional MAC scheme, \eg, HMAC, and shared key $\bar{k}^3_N$. When the controller receives an alert, it determines if id is reported before, if so, it ignores the alert. Otherwise, it sends a request (id) to each node $N$ through an authenticated channel achieved with HMAC and $\bar{k}^3_N$. This request demands each node to report its incoming subspaces.

{\bf 5) Reporting Subspaces.}\quad
Upon receiving the request (id) from the controller, each node $N$ checks if it receives a similar request for the same id before, if it does, it ignores the request. Otherwise, it prepares the report as follows: For each parent node $P$, let $(\vct{y}_1, t_{1,1}, \cdots, t_{1,\lambda}), \cdots, (\vct{y}_l, t_{l,1}, \cdots, t_{l,\lambda})$ be packets of source space id and their tags that node $N$ received from node $P$. Node $N$ sends to the controller through an authenticated channel achieved with HMAC and $\bar{k}^3_N$ the report $(N\,||\,P\,||\,\vct{y}_r\,||\,t_1\,||\,\cdots\,||\,t_\lambda)$, where $\alpha_i \overset{R}{\leftarrow} \mathbb{F}_q$ ($i \in [l]$); $\vct{y}_r = \sum_{i=1}^l \alpha_i \vct{y}_i$; $t_j = \sum_{i=1}^l \alpha_i t_{i,j}$ ($j \in [\lambda]$).

{\bf 6) Locating the Attackers.} After sending out the requests, the controller waits for the reports. After $\Delta t$ seconds, it starts identifying the pollution attackers. It classifies any node that does not report all of its incoming spaces as a malicious node. It only accepts reports with at least $\theta$ valid $\mathsf{SpaceMac}$ tags, where the validation uses keys in $\overline{\mathcal{Y}_{PN}}$~'s. It then identifies the polluted edges in the network based on the reported spaces and the source space. We note that checking if a reported space is polluted can be done quickly and efficiently in $O(mn)$ in terms of multiplication operations by leveraging the global coding coefficients of the reported packet and the source packets. Finally, any node that does not have a polluted incoming edge but has a polluted outgoing edge is classified as malicious.

{\bf 7) Releasing the Result.} After identifying the set of attackers $\mathcal{A}$, the controller sends $(\mathcal{A})$ to each benign node $N$ through an authenticated channel achieved with HMAC and key $\bar{k}^3_N$. Upon receiving $(\mathcal{A})$, each node $N$ adds the nodes in $\mathcal{A}$ into its blacklist. Every node in the network will neither send nor receive traffic from nodes in its blacklist in subsequent communication. The controller itself removes nodes in $\mathcal{A}$ as well as incoming and outgoing edges of these nodes from the network topology. Note that the MAC keys used in the detection scheme do not need to be refreshed when a node is removed from the network. This is because the parent-child relationship between any pair of the remaining adjacent nodes is the same as before the removal.

\section{Security Analysis}
\label{sec:security}

\subsection{Single Adversary}
\label{subsec:single_adversary} 
We described how our our detection and locating scheme work when there is a single adversary when we describe our schemes. We refer the reader to Section \ref{sec:detection} and Section \ref{sec:locating} for the details.

\subsection{Multiple Adversaries}
\label{subsec:multiple_adversaries}

{\bf 1) Detection Scheme:}

{\em a. Independent Adversaries:} We consider adversaries as independent when every adversary checks the integrity of the packets it receives from its parents and drop corrupted packets. This scenario is similar to the single adversary scenario. In-network detection works because if a node $N$ wants to pollute the network, it still has to forge a valid $\SMac$ verification tag for a packet that lies outside of its received space $\Pi_N$, which is computationally difficult.

{\em b. Colluding Adversaries:} We first consider the {\em passive} colluding scenario, where there is an adversary who does not drop corrupted packets that it receives from its parents. In this scenario, in-network detection no longer works. To see this, consider the case where $N$ receives from one of its parents, $P$, a corrupted packet $ \{ t^{k_{\bar{N}}}_{\vct{y}} \,||\,  t^{k_{\bar{P}}}_{\vct{y}} \,||\, \vct{y} \} $ with correct helper tag,  $t^{k_{\bar{N}}}_{\vct{y}}$, and invalid verification tag, $t^{k_{\bar{P}}}_{\vct{y}}$, and $N$ does not drop this packet. To propagate the pollution to its child $C$, it simply computes an appropriate helper tag, $ t^{k_{\bar{C}}}_{\vct{y}}$, for this packet using $k_{\bar{C}}$, then forward the packet $ \{ t^{k_{\bar{C}}}_{\vct{y}} \,||\,  t^{k_{\bar{N}}}_{\vct{y}} \,||\, \vct{y} \} $ to $C$. Clearly, $t^{k_{\bar{N}}}_{\vct{y}}$ is a valid tag for $\vct{y}$; hence, $\vct{y}$ passes $C$ verification, and  $t^{k_{\bar{C}}}_{\vct{y}}$ is a valid helper tag for $\vct{y}$ which will pass any verification by a child of $C$.

We now consider the {\em active} colluding scenario, where a node $N$ can collude with one of its parents, $P$, to learn about the private key, $k_{\bar{N}}$, that is used for verification by its children. In this scenario, in-network detection also fails. This is because knowing the secret key, $k_{\bar{N}}$, $N$ can generate a valid verification tag for any packet outside of its received space $\Pi_N$; thus, any child $C$ of $N$ would not be able to detect corrupted packets sent by $N$.

Note that in both cases where the in-network detection fails, the adversaries must be {\em adjacent} to each other. In both of these cases, the end-to-end detection made by the receivers comes to the rescue. This end-to-end detection is reliable because the receivers are trusted and the private key $k^*$ shared by the source and the receivers is not known to any adversary. In order to pass the verification done by the receivers, an adversary has to forge a valid $\SMac$ tag which is computationally difficult. We discuss how we could relax the assumption of trusted receivers in Section \ref{subsec:mal_receivers}.

{\bf 2) Locating Scheme:}\\
In the presence of multiple adversaries, an attacker may be ``in the shadow'' of some other attackers, which means that it may pollute only already polluted data and thus does not produce any detectable effect. More precisely, we define shadowed and exposed attackers below. 

\begin{defn}[Adapted from \cite{Jafarisiavoshani2008} and \cite{Wang2010}]\label{defn:shadow}
An attacker is \emph{shadowed} if it has at least one polluted incoming edge and is \emph{exposed} otherwise.
\end{defn}


\emph{a. Independent Adversaries:} In this case, we note that with high probability, our approach is already able to identify all exposed attackers. We utilize the following observation to identify all shadowed and exposed attackers.

\begin{lem}\label{lem:oneExposed}
For any directed acyclic graph with pollution attack in presence, there is at least one exposed attacker.
\end{lem}

\begin{IEEEproof}
Consider a topological ordering of the graph, the first malicious node in the ordering is an exposed attacker.
\end{IEEEproof}

Exploiting this, we can use multiple generations, \emph{i.e.}, transmissions of (different) source spaces, to identify all attackers.

\begin{lem}\label{lem:independentDetection}
In a network with $\eta$ independent attackers. With high probability (depending on $q$, $\lambda$, $\delta$, and $\theta$), all attackers can be identified after $\kappa$ generations which experience pollution attack, where $\kappa \le \eta$.
\end{lem}

\begin{IEEEproof}
Since there is at least one exposed attacker per generation by Lemma \ref{lem:oneExposed}, our scheme can identify at least one attacker per generation. Because the identified attackers are immediately excluded from future communication and the other attackers are persistent, subsequent identified attackers are different from the already identified ones. Therefore, it takes at most $\eta$ generations to identify all $\eta$ attackers.
\end{IEEEproof}

Note that we consider the cases where there exists an attacker who is disconnected from the receivers after the removal of all other attackers as degenerate cases. This is because the disconnected attacker is no longer able to pollute the network. In this case, the location of all attackers cannot be determined.

\emph{b. Colluding Adversaries:} We note that each pair of parent $P$ and child $N$ uses distinct key sets $\mathcal{X}_{PN}$ and $\mathcal{Y}_{PN}$; thus,  the collusion of malicious  nodes does not provide knowledge about the key sets of benign nodes. However, when the distance between any two attackers equals to one, where distance refers to the length of the shortest path connecting two nodes, these attackers can collude to report a false space.

\begin{figure}[tp]
\beforeSpace
\centering
\includegraphics[width=8.8cm]{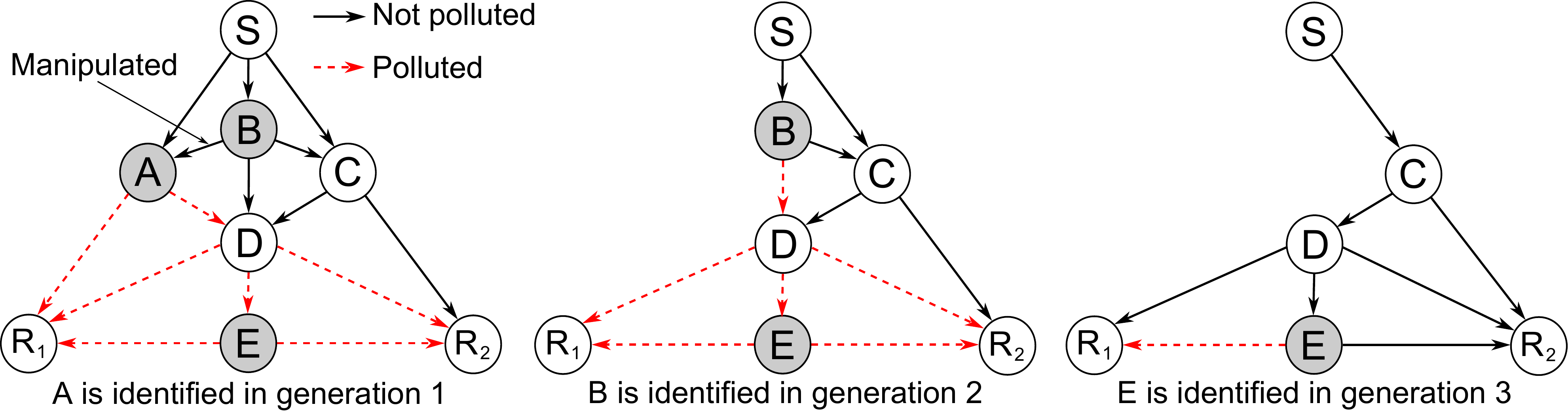}
\vspace*{-10pt}
\caption{An example where there are three attackers $A$, $B$, and $E$. Attackers $A$ and $B$ collude to make edge $e(B,A)$, which is polluted, non-polluted.
Nevertheless, all are identified after 3 generations.}
\label{fig:collude}
\afterSpace
\end{figure}

Assume that in a network, there are colluding attackers $P$ and $N$ connected by a directed edge $e(P, N)$ and there is no other pair of attackers in the network having distance one. We ask the question: ``What can $P$ and $N$ achieve by manipulating edge $e(P,N)$?'' Consider a topological ordering $\mathcal{O}$ of the nodes. If $N$ makes $e(P,N) \in \mathcal{E}_p$, the set of polluted edges identified by the controller, then $P$ will be exposed and identified after all malicious nodes that come before $P$ in $\mathcal{O}$ are identified. After $P$ is located, $N$ and the rest of the attackers will be eventually located. Otherwise, if $N$ makes $e(P,N) \in \mathcal{E}_s$, the set of non-polluted edges identified by the controller, then $N$ will be exposed and located after all malicious nodes that come before $N$ in $\mathcal{O}$ are located. Analogous to the other case, after $N$ is located, $P$ (if not already located) and the rest of the attackers will be eventually located. Consequently, by manipulating the status of edge $e(P,N)$, the attackers can, at best, change the order in which $P$ and $N$ are located. The above analysis can be extended to the general case where there are multiple pairs having distances one by considering the pair $(P,N)$,  where $N$ has a polluted outgoing edge, that appears first in $\mathcal{O}$ first. Fig. \ref{fig:collude} shows an example. As a result, we can generalize lemma \ref{lem:independentDetection}:
\begin{lem}
In a network with $\eta$ attackers. With high probability (depending on $q$, $\lambda$, $\delta$, and $\theta$), all attackers can be identified after $\kappa$ generations which experience pollution attack, where $\kappa \le \eta$.
\end{lem}

\subsection{Tag-Pollution Resistance}
\label{subsec:tag_pollution}
As pointed out by Li \ea~\cite{Li2010}, a scheme that uses multiple MAC tags, such as \cite{Agrawal2009, Yu2009}, may suffer from {\em tag pollution} attacks. In these schemes, a packet carries multiple tags and each node only has keys to verify a subset of them; therefore, an adversary may tamper with some of the tags which only get verified far down the information flow. The consequence is that a packet with some corrupted tags may pass the verification of a few level of nodes. When mixed with other packets, one corrupted tag may snowball into a large number of corrupted tags. The packets carrying these corrupted tags eventually fail the authentication down the stream, thus wasting resources of the network. This effectively emulates a pollution attack.

Both our detection scheme and locating scheme use multiple MAC tags; fortunately, our schemes are resistant to tag pollution attack. More specifically, in our detection scheme, each packet carries three tags: one end-to-end tag, one helper tag, and one verification tag. An adversary cannot tamper with the end-to-end tag because the verification tag of a packet is computed over the concatenation of both the content of the packet and its end-to-end tag. In other words, if the adversary tampers with the end-to-end tag of a packet, the packet will not pass the verification test made by the immediate downstream nodes. This idea of using nested tags was originally introduced in \cite{Li2010}. Apparently, the adversary cannot tamper with the verification tag of a packet since this verification tag is checked immediately by one of its children. Finally, the adversary may tamper with the helper tag of a packet it sends to its child, \eg, attaches an erroneous helper tag. In this case, since the child uses the helper tag to compute verification tags for its outgoing packets, any packet involving the packet with an erroneous helper tag, sent by the child, will have a corrupted verification tag. The next hop that receives packets from this child will drop any packet with a corrupted verification tag. Note that no other tags, besides the end-to-end tags, travel more than two hops in our detection scheme. This eliminates the scenarios where tags are only verified far down the stream as in a tag pollution attack. Our locating scheme uses $\lambda$ tags; however, these tags are never forwarded to any next hop other than the controller (only during the locating process). Furthermore, Lemma \ref{lem:non-reput-sndr} ensures that an adversary can only trick the controller by sending erroneous tags with negligible probability. As a result, our locating scheme is also not susceptible to tag pollution attacks.

\subsection{Denial of Service Attack}
\label{subsec:DoS}
In our locating scheme, once received an alert of pollution from one of the node, the controller triggers the locating process. This involves collecting report vectors from every node in the network. An adversary can exhaust the resource of the network by flooding the controller with alerts. Our locating scheme can combat this denial of service attack in a couple of ways. Recall that the locating process is only triggered once per generation since the controller only issues one request of report per unique generation id. This already limits the effect of the attack. In addition, the controller can maintain a counter, $ctr$, per node. Every time a node $N$ report $(\text{id}\,||\,N)$ and it turns out that there is no polluted edge, $ctr$ is incremented. The controller then ignore any report from $N$ for a period of time if $ctr$ exceeds a certain threshold $\tau$. After this period of time, the counter resets. Note that a malicious node $N$ cannot pretend to be another node $N'$ when sending an alert, \ie, sending $(\text{id}\,||\,N')$ because the alert will be authenticated by the controller using the private key, $\bar{k}^3_{N'}$ shared by the controller and node $N'$.

\subsection{Malicious Receivers}
\label{subsec:mal_receivers}
The assumption that the receivers are trustworthy safeguards our detection scheme against the scenarios where there are adjacent colluding adversaries. 
Here we discuss several available options we could adopt when we relax this assumption and consider the case where some (but not all) receivers are malicious.

One option is to use a separate key for each receiver. When some of the receivers are malicious, it is necessary for the source to share with each receiver, $R_i$, a separate secret key, $k^*_i$, instead of having all receivers and the source share a single key $k^*$. This is because if only one key $k^*$ is used, a malicious receiver can leak the key to an intermediate node; as a result, this node can generate a valid end-to-end tag for any corrupted packet. For this option, for each $k^*_i$, the source generates a different end-to-end $\SMac$ tag; thus, an honest receiver is still able to detect the pollution attack. This approach clearly increases the communication overhead of the end-to-end detection by $| \mathcal{R} |$ times as a packet now carries $| \mathcal{R} |$ end-to-end tags instead of one. This option works when the number of receivers in the network is small. 

Another option is to allow alternative ways of detection. Recall that our location scheme works independently of any detection scheme used. If an honest receiver, $R$, receives a corrupted packet, besides relying on the end-to-end MAC tag to detect corruption, it can also use other knowledge for detection. For instance, as soon as the packets $R$ received form an inconsistent system of equations, $R$ knows there is an attack. Also, $R$ can rely on application-level information to determine corrupted packets. For instance, assume $R$ is able to solve the system but it gets corrupted packets after solving the system, and assume that this is a video packet. The corrupted packet is very likely not compliant with the expected video codec. Using this application-level information, $R$ can detect the pollution as well. As soon as $R$ detects the pollution attack and alerts the controller, the locating process kicks in. Recall that for any generation which experiences a pollution attack, our locating scheme can eliminate at least one attacker. A round of elimination by the locating scheme may break the adjacency property of the attackers, thus enabling the in-network detection to work in the next generation.

\section{Performance Evaluation}
\label{sec:evaluation}

In this section, we evaluate the performance of our detection and locating schemes. We also compare the overhead of our schemes to recently proposed schemes. In addition, we implement $\SMac$ as an open-source library  and we make it available online. Finally, we simulate the scenario when there are multiple adversaries and show that our locating scheme can eliminate all of them within a few generations.

\subsection{Key Management Overhead}
We compare the number of MAC keys that each verifying node needs to maintain in our defense system to that required by state-of-the-art schemes. We start by comparing the overhead of our detection scheme to those of the other MAC-based detection schemes \cite{Li2010, Agrawal2009}. Similar to our detection scheme, the schemes in \cite{Agrawal2009} and \cite{Li2010} also require each node to manage multiple keys. The number of keys could be large in both \cite{Agrawal2009} and \cite{Li2010}. In particular, in \cite{Agrawal2009}, the number of keys each node maintains is, on average, no less than  $\frac{\text{exp}(c+1)^2 \text{ln}(|\mathcal{V}|-1)}{c+1}$, where $c$ is the colluding parameter. Hence, the larger $c$ is, and/or the larger the number of nodes the network has, the more keys are needed. In RIPPLE \cite{Li2010}, keys expire quickly periodically, and new keys are needed frequently; the number of keys increases linearly in the number of time intervals, as the transmission progresses. Clearly, the more time it takes to transmit a generation, and/or the more generations are transmitted, the larger the number of keys is needed. Standing in stark contrast to \cite{Agrawal2009} and \cite{Li2010}, the number of keys a node needs to manage in our detection scheme neither depends on the transmission time nor $c$: it equals to the number of the parents and children the node has (plus one). The number of children and parents of a node may or may not depend on the network size, depending on the network topology. Finally, the number of keys needed for our locating scheme is equal to that of the scheme in \cite{Wang2010}. However, we stress that our keys can be used for multiple generations while this is not the case in \cite{Wang2010} due to replay attacks). 

\subsection{Communication Overhead}

\begin{table*}[t]
\beforeSpace
\centering
\begin{tabular}{|l|c|c|l|c|}
\hline
{\bf Scheme} & {\bf Overhead (bits)} & {\bf In-network Detection} & {\bf Collusion Resistance} & {\bf Tag Pollution Resistance}\\
\hline
Our scheme & $3\, \lceil \text{log}_2 q \rceil$ & Yes & Yes, arbitrary resistance & Yes\\
\hline
RIPPLE \cite{Li2010} & $\frac{\ell}{2} \lceil \text{log}_2 q \rceil$ & Yes & Yes, arbitrary resistance & Yes\\
\hline
Broadcast MAC \cite{Agrawal2009} & $|\mathbb{B}| \lceil \text{log}_2 q \rceil$ & Yes & Yes, $c$ resistance & No\\
\hline
\end{tabular}
\caption{\textnormal{Communication overhead of detection schemes which use homomorphic MACs along with the supported features}}
\label{tab:comm_overhead}
\afterSpace
\end{table*}

Communication overhead refers to the additional network bandwidth that our schemes introduce to the system. For both of the schemes, we neglect the bandwidth of the bootstrapping phase, where symmetric keys are distributed, as this can be done offline.

{\bf Detection scheme.}\quad
For the online overhead per packet, our detection scheme requires each packet to carry three $\SMac$ tags: an end-to-end tag, a helper tag, and a verification tag. Each tag is a symbol in the field $\mathbb{F}_q$; hence, the total overhead is $3\,|q| = 3\, \lceil \text{log}_2 q \rceil$ bits. Our communication overhead is fixed, regardless of the network topology.

Unlike our scheme, the online overhead of the scheme proposed by Li \ea~\cite{Li2010} varies depending on the network level. In particular, in \cite{Li2010}, the authors define a level of a node $N$ as the length of the longest path from $S$ to $N$. The network level $\ell$ is defined as the maximum among the levels of the nodes. In their scheme, each packet carries $\ell$ MAC tags initially, then one or more tags are peeled off at every node the packet goes through. The average overhead is therefore approximately $\frac{\ell}{2} \lceil \text{log}_2 q \rceil$ bits, which is linear in the network level.

In \cite{Agrawal2009}, to achieve security $\frac{1}{q^d}$ and $c$-collusion resistance, \ie, secure against any $c$ colluding attackers, each packet carries $|\mathbb{X}|$ MAC tags and each node verifies $|\mathbb{B}|$ tags, where $(\mathbb{X}, \mathbb{B})$ is a $(c,d)$-cover free family. For instance, to provide security $\frac{1}{q}$ and 2-collusion resistance, each packet needs to carry 49 tags and each node verifies 7 out of these 49 tags \cite{Agrawal2009}. The over head is $49 \lceil \text{log}_2 q \rceil$ in this case, or $|\mathbb{B}| \lceil \text{log}_2 q \rceil$ in general.

Compared to these two schemes, our detection scheme is able to provide in-network detection with significantly less communication overhead because of two main reasons: (i) we exploit local subspace property (Lemma \ref{lem1}) and (ii) we delegate the handling of colluding attackers to our locating scheme. Table \ref{tab:comm_overhead} summarizes the overhead of our detection scheme in comparison to the other two schemes along with the supported features.

\begin{table}[tp]
\centering
\beforeSpace
\begin{tabular}{|c|c|c|c||c|c|c|}
\hline
$q$ & $\lambda$ & $\delta$ & $\theta$ & {\bf Pr}[$P$] & {\bf Pr}[$N$] & {\bf Space Overhead}\\
\hline
$2^8$ & 19 & 9 & 3 & $2^{-10}$ & $2^{-17}$ & 20 bytes\\
$2^8$ & 24 & 12 & 3 & $2^{-14}$ & $2^{-16}$ & 25 bytes\\
$2^8$ & 29 & 14 & 4 & $2^{-16}$ & $2^{-21}$ & 30 bytes\\
\hline
\end{tabular}
\caption{\textnormal{The probability a malicious parent succeeds in preventing its child to report, Pr[$P$], the probability a malicious child succeeds in disparaging its parent, Pr[$N$], and the space overhead correspond to different parameter sets.}}
\label{table:params}
\afterSpace
\vspace*{-15pt}
\end{table}

{\bf Locating scheme.}\quad
The communication overhead of our locating scheme includes the $\lambda$ tags carried by each packet, the reporting vectors sent by the nodes, and the announcement (containing the identified adversaries) sent by the controller. The online overhead, which depends on the number of packets sent in the network is the first one; the latter two overhead exist only when there is a pollution attack detected, thus are asymptotically negligible in the number of packets. The overhead per packet is $\lambda\,\lceil \text{log}_2 q \rceil$ bits. Table \ref{table:params} shows that with an overhead of about 20 bytes per packet, the probability that a malicious parent succeeds in preventing its child to report, Pr[$P$], and the probability that a malicious child succeeds in disparaging its parent, Pr[$N$], are both very small.


Compared to the scheme by Wang \ea~\cite{Wang2010}, we have the same amount of online overhead per packet ($\lambda$ tags). However, in \cite{Wang2010}, when a pollution attack is detected, the controller has to compute multiple checksums for the polluted generation and send these checksums to all the nodes. Each checksum includes $m$ symbols in $F_q$ (recall that $m$ is the number of packets per generation). If $\mu$ checksums are computed ($\mu>1$ to improve the security guarantee), the overhead resulted in sending the checksums to all nodes is $|\mathcal{V}| \, \mu \, m \, \lceil \text{log}_2 q \rceil$ bits. In contrast, our locating scheme does not need this checksum dissemination.

{\bf Combined Scheme.}\quad
The total online communication overhead of our defense scheme is $(3+\lambda)\,\lceil \text{log}_2 q \rceil$ bits. Note that this overhead neither depends on the packet size nor the generation size; hence, it becomes more inexpensive when the packet size is large. For instance, for $\lambda=19, n=1024, m=32, q=2^8$, the per-packet communication overhead is only 2\% while the security of the detection is $2^{-8}$ and of the security of the locating is $2^{-10}$. Most importantly, standing in stark contrast to the other two schemes: \cite{Agrawal2009} and \cite{Li2010}, our overhead is constant in terms of the network size and the number of attackers.

\subsection{Computation Overhead}

The major computational overhead of both of our schemes are from the algorithms $\mac$, $\cbn$, and $\vrf$ performed at each node for every packet. The computation cost for the bootstrapping, reporting, and locating steps is a one-time cost for a generation and thus is asymptotically negligible in the number of packets. We subsequently focus on the online computation cost incurred by the three algorithms of $\SMac$.

Both the $\mac$ and the $\vrf$ algorithms incur one PRG call, $m$ PRF calls, and $(n+2m)$ finite field multiplications. Note that the results of both the PRG and PRF calls can be cached and used for the whole generation. Thus, they can be considered as a one-time cost as well. If we let $w$ be the average number of packets combined by each node, then the $\cbn$ algorithm incurs $w$ multiplications on average. 

{\bf Detection Scheme.}\quad The operations performed by each node in the detection scheme for each packet $\vct{y}$ include (i) verifying the integrity of $\vct{y}$ using the $\vrf$ algorithm, (ii) combining the received helper tags to generate a verification tag for an outgoing packet $\vct{z}$ using the $\cbn$ algorithm, (iii) computing a helper tag for $\vct{z}$ using the $\mac$ algorithm. If the node is the source, it needs to compute the end-to-end tag using the $\mac$ algorithm; however, in this case it does not need to verify the integrity of packets. If it is a receiver, it needs to verify the end-to-end tag by performing another $\vrf$ algorithm. The worst case computational overhead, \ie, when the node is a receiver, in terms of the number of finite field multiplications is $3\,(n+2m) + w$.

{\bf Locating Scheme.}\quad For each packet received, each node verifies $\delta$ tags using the $\vrf$ algorithm. For each packet it sends out, each node needs to compute $\lambda$ tags using the $\mac$ algorithm. The total overhead is therefore  $(\delta+\lambda)\,(n+2m)$ number of multiplications.

{\bf Combined Scheme.}\quad The overall computational overhead per packet per node of our combined scheme is 
$(3+\delta+\lambda)(n+2m) + w\,.$

{\bf Comparison.}\quad The computational overhead per node per packet of the scheme proposed by Li \ea~\cite{Li2010} includes one $\cbn$ and one $\vrf$ operation. Based on the closed-form formulas provided in \cite{Li2010}, this overhead is
$w\,(\frac{\ell-1}{2}) + (n+m+\frac{\ell-1}{2})$ number of multiplications.
Finally, the computational overhead per node per packet of the scheme proposed by Agrawal and Boneh \cite{Agrawal2009} includes one $\cbn$ and $|\mathbb{B}|$ $\vrf$, which is
$w\,|\mathbb{X}| + |\mathbb{B}|\,(n+2m)$ number of multiplications.

For a concrete comparison, let $q=2^8, n=1024, m=32, w=4, \ell=9$ (for an average network of size 100, note that $\text{log}_2 100 \approx 6$). Table \ref{tab:computation_overhead} shows the set of appropriate parameters and their corresponding computation overhead to achieve the security $2^{-8}$ for all schemes. Following \cite{Agrawal2009}, we implement multiplication in $\mathbb{F}_{2^8}$ by creating an offline multiplication table, storing all $2^{16}$ products of pairs of elements in this field. The table only occupies about 64 KB in C/C++ and 128 KB in Java (since there is no 8-bit primitive data type in Java that has values from 0 to $2^8$-1). This table enables us to achieve fast multiplication, which is now just a table lookup. Note that this approach is not possible (space-wise) when working with large field, for instance, any scheme that relies on public key cryptography, such as, \cite{Boneh2009, Charles2006, Zhao2007}, requires a large field size, \ie, $\lceil \text{log}_2 q \rceil \geq 128$. The details about the platforms we use are provided in Section \ref{subsec:library}. The numbers reported in Table \ref{tab:computation_overhead} are averaged over $10^6$ multiplications.

As shown in Table \ref{tab:computation_overhead}, the computational latency of our detection scheme is in the same order of magnitude as the other two detection schemes. 
Table \ref{tab:computation_overhead} also shows that the latency of our combined detection-locating scheme is about 10 times higher than that of our detection scheme, or 30 times higher than that of RIPPLE.

This is the trade-off when one wants to locate and eliminate all attackers. If one chooses not to locate and eliminate the attackers, they may keep flooding their child nodes with corrupted packets along with their MAC tags. This not only wastes the child nodes' download bandwidth, but also exhausts their computational resources since they need to constantly run the verification algorithm on the corrupted packets. Furthermore, the attackers may only send out corrupted packets but not valid packets. This means that all packets they receive from their parent nodes are not used at all, which implies that the upload bandwidth of the parent nodes is also wasted. Apparently, the more parents and children the attackers have, and/or the larger the number of attackers the network has, the more resources are wasted due to the attack.

Nevertheless, we note that even though the combined detection-locating scheme has one order of magnitude larger computational delay than other stand-alone detection schemes, its delay is still very small, in the order of sub-millisecond on a PC or millisecond on a resource-constrained Android phone (Samsung Captivate). Therefore, when operating on PCs or smart phones, we strongly recommend using our full scheme. In  scenarios where the network cannot afford the computation overhead of the locating scheme, each node may want to keep a threshold (per parent) of how many corrupted packets it detects from this parent so far, and refuse to receive packets from the parent after the number of corrupted packets crosses this threshold. This reduces the waste of the node's download bandwidth and CPU time.

\begin{table*}[t]
\beforeSpace
\centering
\begin{tabular}{|l|l|l|l|l|}
\hline
{~} & {\bf RIPPLE} \cite{Li2010} & {\bf Broadcast MAC} \cite{Agrawal2009} & {\bf Our Detection} & {\bf Our Full System}\\
\hline
\multirow{4}{*}{\bf Features} & In-network detection & In-network detection & In-network detection & In-network detection\\
   & Arbitrary collusion res. & $c$-collusion res. & Arbitrary collusion res. & Arbitrary collusion res.\\
   & Tag-pollution res. & {\em No} tag pollution res. & Tag pollution res. & Tag pollution res.\\
   & {\em No} locating & {\em No} locating & {\em No} locating & Exact locating\\
\hline
{\bf Network Params.} & \multicolumn{4}{|c|}{$q=2^8, n=1024, m=32, w=4, \ell=9$}\\
\hline
{\bf Scheme Params.} & ~ & $|\mathbb{X}|=49, |\mathbb{B}|=7, c=2$ & ~ & $\lambda=19, \delta=9, \theta=3$\\
\hline
{\bf Security} & \multicolumn{4}{|c|}{$2^{-8}$}\\
\hline
\hline
{\bf \# Multiplications} & 1,096 & 7,812 & 3,268 & 33,732\\
\hline
{\bf C/C++} ($\mu$s) & 5.5 & 39.1 & 16.3 & 168.7\\ 
\hline
{\bf Java} ($\mu$s) & 6.6 & 46.9 & 19.6 & 202.4\\ 
\hline
{\bf Android} ($\mu$s) & 116.2 & 828.1 & 346.4 & 3,575.6\\ 
\hline
\end{tabular}
\caption{\textnormal{Online computation overhead per packet per node in terms of the number of finite field multiplications, computing latency in C/C++, Java, and on an Android platform (Samsung Captivate)}}
\label{tab:computation_overhead}
\afterSpace
\end{table*}


\subsection{Library}
\label{subsec:library}

\begin{table}[t]
\beforeSpace
\centering
\begin{tabular}{|c|c|l|l|l|l|}
\hline
~ & {\bf \# Tags} & $\mac$ & $\cbn$ & $\vrf$ & {\bf Security}\\
\hline
\multirow{2}{*}{C/C++} & 1 & 28 & 0.02 & 28 & $2^{-8}$\\
   \cline{2-6}
   & 4 & 112 & 0.08 & 112 & $2^{-32}$\\
\hline
\multirow{2}{*}{Java} & 1 & 61 & 0.09 & 61 & $2^{-8}$\\
   \cline{2-6}
   & 4 & 244 & 0.36 & 244 & $2^{-32}$\\
\hline
\multirow{2}{*}{Android} & 1 & 2,273 & 0.66 & 2,273 & $2^{-8}$\\
   \cline{2-6}
   & 4 & 9,092 & 2.64 & 9,092 & $2^{-32}$\\
\hline
\end{tabular}
\caption{\textnormal{Computational time in $\mu$s of $\SMac$ algorithms in $\mathbb{F}_{2^8}$}}
\label{tab:benchmark}
\afterSpace
\end{table}

We implement all three algorithms of $\SMac$ in both C/C++ and Java and provide them as a library. As mentioned before, we implement field multiplication using a look-up table. We implement addition as a simple XOR operation. Finally, we implement PRF and PRG using AES with CBC mode of operation. The AES implementation is provided by the standard {\em crypto} library \cite{JavaxCrypto} for Java implementation and {\em crypto++} open-source library \cite{CryptoPP} for C/C++ implementation. We make our $\SMac$ library available online along with the source code \cite{SpaceMacLib}.

This library is useful for those who want to adopt our $\SMac$ scheme into their system, or those who want to deploy our proposed defense scheme. The C/C++ implementation is faster than the Java implementation; it is meant to be used by low-level or embedded devices, such as, network routers. The Java implementation, meanwhile, is useful for high-level application-layer programs, such as, peer-to-peer applications. Furthermore, the Java implementation is ready to be run on the current Android OS (Android 2.2 Froyo). This provides support for the rising implementation of network coding on smart phones, such as, the work in \cite{Fitzek2010}, \cite{Shojania2009}, and \cite{Shojania2010}.

Table \ref{tab:benchmark} provides the benchmark of all three algorithms of $\SMac$. For the benchmark, we set $n=1024, m=32, w=4$. Except for the Android benchmark, both the C/C++ and Java implementations were run on a PC with a quad-core 2.8 Ghz processor and 32 GB of RAM. Our Android device, the Samsung Captivate, has a single 1 Ghz processor and 512 MB RAM. The reported values correspond to the averages taken over $10^5$ runs of each algorithm. The most expensive operations of $\mac$ and $\vrf$ algorithms are the PRF and PRG calls; however, we stress that these calls can be done offline. For completeness, the reported values include the cost of these calls.

From Table \ref{tab:benchmark}, we can see that in order to achieve high security ($2^{-32}$), the computational latency of our C/C++ implementation is only in the order of hundreds of microseconds. Moreover, even on the Android resource-constrained device, the computational latency of $\mac$ and $\vrf$ are still very small, only in the order of millisecond. Note that the $\cbn$ operation is several orders of magnitude faster than the $\mac$ and $\vrf$ algorithms since it only involves finite field multiplications, which are quick table lookups. These results demonstrate that $\SMac$ algorithms are fast and appropriate for practical use.

\subsection{Simulation}
\begin{table}[tp]
\beforeSpace
\centering
\begin{tabular}{|l|c|c|c|c|c|}
\hline
{\bf Number of attackers} & 4 & 8 & 12 & 16 & 20\\
\hline
{\bf Average \# of generations} & 1.92 & 3.01 & 4.16 & 4.77 & 5.64\\
\hline
{\bf Average delay} (ms) & 412 & 647 & 896 & 1,031 & 1,217\\
\hline
\end{tabular}
\caption{\textnormal{The average number of generations and delay required to detect all attackers in a network with 50 intermediate nodes.}}
\label{table:locatingDelay}
\afterSpace
\end{table}

We implement a simulation in Python that simulates a scenario where there are multiple colluding attackers in a network. We generate between a pair of source and receiver nodes a random directed acyclic graph network of 50 nodes using the {\em pygraph} library \cite{PedroMatiello}. The ratio of edges to nodes is a random number in $[1,5]$. All edges have a random end-to-end delay between 10 and 100 ms. In a single generation, 32 packets in $\mathbb{F}_{2^8}^{1056}$ ($q=2^8, n=1024, m=32$) are generated and sent by the source node. The locating process is triggered multiple times, each time by an alert by the receiver.

The attackers in the network are chosen randomly from the population of 50 nodes in the network in a way that each attacker can still pollute the network even when the rest attackers are removed. The attackers pollute all of their outgoing edges. When requested by the controller, most of them honestly report their incoming subspaces; however, some of them, who have malicious parents, lie about their received subspaces from those parents. This emulates the case where the attackers collude to manipulate the reports of their incoming subspaces.

We evaluate the average number of generations to locate all $\eta$ attackers in the network, where $\eta$ varies from 4 to 20. For each $\eta$, we perform the simulation for 100 rounds (varying the network topology, attacker location, and edge delays) to get the average value. We also evaluate the average delay it takes to identify all attackers, where the delay refers to the time between when the source starts sending and when all the attackers are located. The results shown in table \ref{table:locatingDelay} indicate that we succeed in locating all attackers very quickly (about second for 20 attackers) and after much smaller than $\eta$ generations (5.5 generations for 20 attackers).

\section{Conclusion}
\label{sec:conclusion}
In this work, we introduce a novel homomorphic MAC scheme for expanding space called $\SMac$. We propose a cooperative defense system against pollution attacks built on $\SMac$. To the best of our knowledge, our system is the first that can provide both in-network detection and exact locating of the attackers. In addition, our system is collusion resistant and tag-pollution resistant. Our evaluation results using real implementation in C/C++ and Java on multiple devices demonstrate that our defense scheme incurs both low communication and low computation overhead. We implemented $\SMac$ as a ready-to-use library and make it available online.



\bibliographystyle{IEEEtran}
\bibliography{NetworkCodingSecurity}






\appendix

\section*{Proof of Lemma \ref{lem:non-reput-rcvr}} 
Recall that the server checks $\lambda - \delta$ tags, and in order for the server to accept a report, there must be at least $\theta$ valid tags. The probability that the child successfully forges a $\mathsf{SpaceMac}$ tag is $\frac{1}{q}$; and so, the probability that the child fails to forge such a tag is $1-\frac{1}{q}$. Let $i$ be the number of valid tags. The stated probability is a direct result of enumerating the probability of success of the child in all the cases.

\section*{Proof of Lemma \ref{lem:non-reput-sndr}} 
Let $\vct{y}_r$ denote the random packet of the parent's space that the child chooses to report to the controller: $\vct{y}_r = \sum_{j \in \mathcal{D}} \alpha_j\,\vct{y}_j$, where $\alpha_j \neq 0$ and $\mathcal{D}$ is a subset of indices of the packets sent from the parent to the child.

Recall that the child is benign and always uses $\mathsf{Combine}$ to generate tag for  $\vct{y}_r$. Let $x$ denote the number of correctly computed tags of $\vct{y}_r$, \emph{i.e.}, the parent uses $\mathsf{Mac}$ to compute the corresponding $x$ tags for every $\vct{y}_j$ for $j \in \mathcal{D}$. The value of $x$ must be smaller than $\delta + \theta$ otherwise the controller will accept $\vct{y}_r$ as there are at least $\theta$ valid tags. Let $i$ out of these $x$ tags be the number of tags verifiable by the child, $i \leq \text{min}(\delta,x)$. Since $\vct{y}_r$ has $\delta - i$ not correctly computed and verifiable by the child, there are at least $\delta-i$ not correctly computed tags which are verifiable by the child among the tags of $\vct{y}_j$'s. Thus, the probability that the child accepts all $\vct{y}_j$'s is at most $\frac{1}{q^{\delta-i}}$. 

The remaining $x - i$ tags of $\vct{y}_r$ are checked by the controller. These tags of $\vct{y}_r$ are valid tags; hence, $x - i$ must be smaller than $\theta$ otherwise the controller will accept the report. Thus, $x-i<\theta$; hence, $i \geq \text{max}(x-\theta+1, 0)$. The probability that the controller rejects $\vct{y}_r$ equals to the probability that there are less than $\theta$ valid tags.  Since there are already $x-i$ valid tags, this probability equals
\[p_{\text{rej}} = \sum_{j=0}^{\theta - (x-i) - 1} \binom{\theta - (x-i) - 1}{j} \frac{1}{q^j} (1 - \frac{1}{q} )^{\lambda - \delta - (x-i) - j}\,,\]
where $j$ denotes the number of valid tags out of the rest $\lambda - \delta - (x-i)$ tags. Note that $p_{\text{rej}} \leq 1$.

Putting the above values together, the probability that the child accepts all $\vct{y}_j$'s which form $\vct{y}_r$ and the server rejects $\vct{y}_r$ when there are $x$ correctly computed tags is upper bounded as follows:
\begin{align*}
p(x) &\leq \frac{1}{\binom{\lambda}{x}} \cdot \sum_{i= \text{max}(x - \theta + 1,0)}^{\text{min}(\delta,x)} \left[\binom{\delta}{i}  \, \binom{\lambda-\delta}{x-i} \cdot \frac{1 }{ q^{\delta - i}} \cdot p_{\text{rej}} \right]\\
~&\leq \frac{1}{\binom{\lambda}{x}} \cdot \sum_{i= \text{max}(x - \theta + 1,0)}^{\text{min}(\delta,x)} \left[ \binom{\delta}{i} \, \binom{\lambda-\delta}{x-i} \cdot \frac{1 }{ q^{\delta - i}} \right]
\end{align*}

The best probability is the maximum of $p(x)$'s.

\balance


%



\end{document}